\documentstyle[prl,aps,epsf]{revtex}

 \newcommand{\insertplot}[5]{\begin{figure}
 \hfill\hbox to 0.05in{\vbox to #5in{\vfill
 \inputplot{#1}{#4}{#5}}\hfill}
 \hfill\vspace{-.1in}
 \caption{#2}\label{#3}
 \end{figure}}
 \newcommand{\inputplot}[3]{
 \special{ps: plotfile #1}
}

\begin{document}
\draft
\title{SEQUENCES OF GLOBALLY REGULAR AND BLACK HOLE SOLUTIONS 
      IN SU(4) EINSTEIN-YANG-MILLS THEORY}
\vspace{1.5truecm}
\author{
{\bf 
Burkhard Kleihaus, Jutta Kunz, Abha Sood, Marion Wirschins}\\
Fachbereich Physik, Universit\"at Oldenburg, Postfach 2503\\
D-26111 Oldenburg, Germany}

\date{\today}

\maketitle
\vspace{1.0truecm}

\begin{abstract}
SU(4) Einstein-Yang-Mills theory possesses sequences of 
static spherically symmetric
globally regular and black hole solutions.
Considering solutions with a purely magnetic gauge field,
based on the 4-dimensional embedding
of $su(2)$ in $su(4)$, these solutions are labelled by the
node numbers $(n_1,n_2,n_3)$ of the three gauge field functions
$u_1$, $u_2$ and $u_3$.
We classify the various types of solutions in sequences and 
determine their limiting solutions.
The limiting solutions of the sequences of neutral solutions
carry charge, and
the limiting solutions of the sequences of charged solutions
carry higher charge.
For sequences of black hole solutions with node structure $(n,j,n)$ and
$(n,n,n)$, several distinct branches of solutions exist
up to critical values of the horizon radius.
We determine the critical behaviour for these sequences of solutions.
We also consider SU(4) Einstein-Yang-Mills-dilaton theory
and show that these sequences of solutions
are analogous in most respects to the corresponding
SU(4) Einstein-Yang-Mills sequences of solutions.
\end{abstract}

\vfill
\noindent {Preprint hep-th/9802143} \hfill\break
\vfill\eject

\section{Introduction}

SU(2) Einstein-Yang-Mills (EYM) theory possesses a sequence
of asymptotically flat static spherically symmetric solutions,
which are globally regular \cite{bm}.
Based on a purely magnetic gauge field ansatz,
these solutions carry no charge.
They are characterized by the node number $n$
of the single magnetic gauge field function $u$.
With increasing node number $n$, the sequence of neutral
solutions tends to a limiting solution
with magnetic charge $P=1$,
which consists of two parts, an inner oscillating part and an outer
Reissner-Nordstr\o m (RN) part
\cite{bfm}.

Besides the globally regular solutions,
SU(2) EYM theory possesses static spherically symmetric
black hole solutions
with non-trivial non-abelian gauge field configurations
outside their regular event horizon \cite{su2}.
For any value of the event horizon radius $x_{\rm H}$,
there exists a corresponding sequence of 
black hole solutions, characterized by the node number $n$
of the gauge field function.
For fixed horizon radius $x_{\rm H}$ and increasing $n$, 
the sequence of neutral black hole solutions also tends
to a limiting charged solution with magnetic charge $P=1$ \cite{bfm}.
For horizon radius $x_{\rm H}>1$, this limiting solution corresponds
to an embedded RN solution with the same
horizon radius $x_{\rm H}$.
For $0<x_{\rm H}<1$, the limiting solution again
consists of an oscillating part and a RN part \cite{bfm}.
Interestingly, the only charged black hole solutions of SU(2) EYM theory
are embedded RN solutions \cite{no-hair}.

In SU(3) EYM theory,
besides embedded RN solutions \cite{yasskin}
charged static spherically symmetric black hole solutions exist,
which possess non-trivial gauge field configurations outside
their regular event horizon \cite{gv,kks4}. 
These black hole solutions have magnetic charge of norm $P=\sqrt{3}$
residing in the $su(3)$ Cartan subalgebra \cite{kks5}.
Like the RN solutions, the magnetically 
charged non-abelian black hole solutions
exist only for horizon radius $x_{\rm H} \ge P$, 
where $x_{\rm H} = P$ leads to extremal black hole solutions
\cite{foot1}.
These charged non-abelian SU(3) black hole solutions are characterized 
by the node number $n$ of a single magnetic gauge field function,
similar to the neutral non-abelian SU(2) black hole solutions.
With increasing $n$ and fixed horizon radius $x_{\rm H}$,
the sequence of non-abelian black hole
solutions with charge $P=\sqrt{3}$ tends to a solution with higher
charge, $P=2$ \cite{kks4}.
For horizon radius $x_{\rm H}>2$, this limiting solution again corresponds
to an embedded RN solution with the same
horizon radius $x_{\rm H}$, whereas
for $\sqrt{3} <x_{\rm H}<2$ the limiting solution again
consists of an oscillating part and a RN part \cite{kks4}.

SU(3) EYM theory also possesses neutral globally regular
and black hole solutions.
The genuine neutral static spherically symmetric SU(3) EYM solutions
are obtained by embedding the 3-dimensional representation
of $su(2)$ in $su(3)$ \cite{kuenzle,kks2,kks4}.
The purely magnetic gauge field ansatz then involves two functions,
$u_1$ and $u_2$,
and the solutions are labelled by the corresponding node numbers
$(n_1,n_2)$.
For fixed horizon radius $x_{\rm H}$, a discrete set of regular 
neutral SU(3) EYM solutions is obtained \cite{foot2}.
These neutral solutions form sequences with node structure $(n,n)$ and
$(i,i+n)$, with $i$ fixed.
With increasing $n$, these sequences of solutions again tend to 
limiting solutions, carrying magnetic charge of norm $P=2$ and
$P=\sqrt{3}$, respectively.
\cite{kks4,kks2}.
For the $(i,i+n)$ sequences, 
the limiting solution of black hole solutions 
with horizon radius $x_{\rm H}>\sqrt{3}$ corresponds
to a charged non-abelian SU(3) black hole solution with $i$ nodes
and with the same horizon radius $x_{\rm H}$, whereas
for black hole solutions with horizon radius 
$0 <x_{\rm H}< \sqrt{3}$ and for globally regular
solutions the limiting solution again
consists of two parts \cite{kks4}.
The $(n,n)$ sequences either represent genuine SU(3) solutions
with $u_1\ne u_2$ or scaled SU(2) solutions with $u_1 = u_2$.
Their limiting solutions carry magnetic charge of norm $P=2$,
and represent embedded RN solutions for $x_{\rm H} > 2$.
Unlike the $(i,i+n)$ sequences of black hole solutions,
the genuine SU(3) $(n,n)$ black hole solutions exist only for 
sufficiently small values of the horizon radius $x_{\rm H}$.
At critical values of the horizon radius, they merge into
the scaled SU(2) solutions. Therefore, of the two 
types of $(n,n)$ sequences  
only the scaled SU(2) solutions persist for large $x_{\rm H}$ \cite{kks4}.

Here we investigate the sequences of SU(4) EYM solutions.
The genuine static spherically symmetric SU(4) solutions
are obtained by embedding the 4-dimensional
representation of $su(2)$ in $su(4)$ \cite{kuenzle,yves}.
The purely magnetic gauge field ansatz then contains three functions,
$u_1$, $u_2$ and $u_3$. The classification of the 
discrete set of regular solutions
in sequences again involves their node numbers $(n_1,n_2,n_3)$
\cite{foot2}.
When some of the gauge field functions possess the same number of nodes,
several solutions can exist for 
sufficiently small values of the horizon radius $x_{\rm H}$,
leading to a complex critical behaviour of the solutions
as functions of $x_{\rm H}$.
When all three gauge field functions are non-trivial, neutral solutions
are obtained, whereas when one or more gauge field functions
are identically zero, charged solutions arise.
We determine the limiting solutions of the sequences of 
globally regular and black hole solutions
and specify their magnetic charge.
The extension to SU(N) EYM theory is straightforward.

By including a dilaton field, all these solutions can be generalized
to the solutions of Einstein-Yang-Mills-dilaton (EYMD) theory \cite{eymd,gal,kks3,kks4}.
The classification of the solutions of EYMD theory and EYM theory
is identical. However, the charged EYMD black hole solutions
exist for arbitrarily small values of the horizon radius $x_{\rm H}$,
and the limiting solutions of some EYMD sequences of solutions
are embedded Einstein-Maxwell-dilaton (EMD) solutions \cite{emd}.

In this paper we present in section II the SU(4) EYM action and the static
spherically symmetric ansatz for the metric and the 
purely magnetic gauge field, based on the 4-dimensional embedding
of $su(2)$ in $su(4)$, and we determine the boundary conditions for
asymptotically flat globally regular and black hole solutions.
In section III we discuss the charged SU(4) EYM black hole solutions.
We classify them and construct numerically several sequences
of charged black hole solutions. In particular, we demonstrate
the exponential convergence of some of their properties.
In section IV we classify the globally regular
SU(4) EYM solutions in sequences.
We consider several sequences in detail
and present numerical results for them.
The sequences of neutral SU(4) EYM black hole solutions
are discussed in section V, where we consider in particular 
the complex pattern of bifurcations,
occurring at critical values of the horizon radius.
In section VI we briefly generalize our results to SU(4) EYMD theory
and present some numerical examples.
We give our conclusions in section VII.

\section{SU(4) Einstein-Yang-Mills Equations of Motion}

\subsection{SU(4) Einstein-Yang-Mills action}

We consider $SU(4)$ EYM theory with action
\begin{equation}
S=S_G+S_M=\int L_G \sqrt{-g} d^4x + \int L_M \sqrt{-g} d^4x
\ ,  \label{action}  \end{equation}
\begin{equation}
L_G=\frac{1}{16\pi G}R
\ , \label{LG} \end{equation}
gauge field Lagrangian
\begin{equation}
L_M=-\frac{1}{2} {\rm Tr} (F_{\mu\nu} F^{\mu\nu})
\ , \label{lagm} \end{equation}
field strength tensor
\begin{equation}
F_{\mu\nu}= \partial_\mu A_\nu - \partial_\nu A_\mu
            - i e [A_\mu,A_\nu]
\ , \label{FMN} \end{equation}
gauge field
\begin{equation}
A_\mu = \frac{1}{2} \lambda^a A_\mu^a
\ , \end{equation}
and gauge coupling constant $e$.

Variation of the action eq.~(\ref{action}) with respect to the metric
$g^{\mu\nu}$ leads to the Einstein equations
\begin{equation}
G_{\mu\nu}= R_{\mu\nu}-\frac{1}{2}g_{\mu\nu}R = 8\pi G T_{\mu\nu}
\ , \end{equation}
with stress-energy tensor
\begin{equation}
T_{\mu\nu}  =  g_{\mu\nu}L_M -2 \frac{\partial L_M}{\partial g^{\mu\nu}} 
   =  2  {\rm Tr} 
    ( F_{\mu\alpha} F_{\nu\beta} g^{\alpha\beta}
   -\frac{1}{4} g_{\mu\nu} F_{\alpha\beta} F^{\alpha\beta})
\ , \end{equation}
and variation with respect to the gauge field $A_\mu$ 
leads to the gauge field equations.

\subsection{Static spherically symmetric ans\"atze}

To construct static spherically symmetric solutions
we employ Schwarz\-schild-like coordinates and adopt
the static spherically symmetric metric
\begin{equation}
ds^2=g_{\mu\nu}dx^\mu dx^\nu=
  - {\cal A}^2{\cal N} dt^2 + {\cal N}^{-1} dr^2 
  + r^2 (d\theta^2 + \sin^2\theta d\phi^2)
\ , \label{metric} \end{equation}
with the metric functions ${\cal A}(r)$ and
\begin{equation}
{\cal N}(r)=1-\frac{2m(r)}{r}
\ . \label{n} \end{equation}

The static spherically symmetric ans\"atze
for the gauge field $A_{\mu}$ of SU(4) EYM theory
are based on the $su(2)$ subalgebras of $su(4)$.
Here we consider the gauge field ansatz corresponding to the $4$-dimensional 
embedding of $su(2)$ in $su(4)$ \cite{kuenzle}
\begin{equation} 
 A_{\mu}  dx^\mu   =  \frac{1}{2e} \left( 
\begin{array}{cccc}
3\cos\theta d\phi & \omega_1 \Theta & 0 & 0 \\
\omega_1 \bar \Theta & \cos\theta d\phi & \omega_2 \Theta & 0 \\
0 & \omega_2 \bar \Theta & -\cos\theta d\phi & \omega_3 \Theta \\
0 & 0 & \omega_3 \bar \Theta & -3\cos\theta d\phi
\   \end{array} \right)
\ , \label{amu} \end{equation}   
with 
\begin{equation}
\Theta = i d \theta + \sin \theta d \phi
\ , \end{equation}   
i.~e.
\begin{equation}
A_0=A_r=0
\ . \end{equation}   
The ansatz contains three gauge field functions
$\omega_j(r)$, $j=1,2,3$,
and leads to the field strength tensor components
\begin{equation} 
F_{r\theta}=\partial_r A_\theta
\ , \end{equation}   
\begin{equation} 
F_{r\phi}=\partial_r A_\phi
\ , \end{equation}   
and
\begin{equation} \label{F1}
F_{\theta\phi}=(1/2e) {\rm diag} (f_1,...,f_4) \sin \theta
\ , \end{equation}   
with
\begin{equation} 
f_j = \omega_j^2 - \omega^2_{j-1} + \delta_j \ , \ \ \
\delta_j = 2j - 5 \ , \ \ \ j=1,\ldots,4 
\  \ \ \ (\omega_0=\omega_4=0)
\ . \label{f1} \end{equation}  
Another parametrization of the gauge field ansatz is given in \cite{yves}.

\subsection{Field equations}

With the above ans\"atze we derive the set of EYM equations.
The metric (\ref{metric}) yields for the $tt$ and $rr$ components of
the Einstein equations
\begin{equation}
G_{tt}=\frac{2m'{\cal A}^2{\cal N}}{r^2} =8\pi G T_{tt}
\ , \label{tt} \end{equation}
and
\begin{equation}
G_{rr} =  -\frac{G_{tt}}{{\cal A}^2{\cal N}^2}
  +\frac{2}{r} \frac{{\cal A}'}{{\cal A}} =8\pi G T_{rr}
\ , \label{rr} \end{equation}
where the prime indicates the derivative with respect to $r$.
The static spherically symmetric ansatz for the fields (\ref{amu})
yields for the $tt$ component of the
stress-energy tensor $T_{tt}=-{\cal A}^2{\cal N}L_M$,
\begin{equation}
 T_{tt} =  \frac{1}{e^2 r^2} {\cal A}^2{\cal N} \left( 
  {{\cal N} {\cal G} + {\cal P}} \right)
\ , \end{equation}
and for the $rr$ component
\begin{equation}
 T_{rr} =  \frac{1}{e^2 r^2 {\cal N}} \left(
  {{\cal N} {\cal G} - {\cal P}} \right)
\ , \end{equation}
with
\begin{equation}
{\cal G} = \sum_{j=1}^{3} \omega_j^{' 2}\ , \ \ \
{\cal P} = \frac{1}{4 r^2} \sum_{j=1}^4 f_j^2
\ . \label{gp} \end{equation}

We introduce the dimensionless coordinate 
\begin{equation}
x=\frac{er}{\sqrt{4\pi G}} 
\  , \label{x} \end{equation}
the dimensionless mass function
\begin{equation}
\mu=\frac{em}{\sqrt{4\pi G}} 
\ , \label{mu} \end{equation}
and the scaled gauge field functions \cite{kuenzle}
\begin{equation}
 u_j = \frac{\omega_j}{\sqrt{\gamma_j}} \ , \ \ \
 \gamma_j = {j (4 - j) } \ , \ \ \ j=1,\ldots,3
\  . \end{equation}
The above Einstein equations then yield for the metric functions
the equations
\begin{eqnarray}
\mu^{'} & = & {\cal N}\cal{G}  + \cal{P}  \quad , \label{sunm}\\
\frac{{\cal A}^{'}}{{\cal A}} & = & \frac{2 {\cal G}}{x} 
\ , \label{suna} \end{eqnarray}
where the prime now indicates the derivative with respect to $x$
and
\begin{equation}
{\cal G} = \sum_{j=1}^{3} \gamma_j u_j^{' 2} \ , \ \ \
{\cal P} = \frac{1}{4 x^2} \sum_{j=1}^4 f_j^2
\ , \label{gp2} \end{equation}
with $f_j = \gamma_j u_j^2 - \gamma_{j-1} u^2_{j-1} + \delta_j$
(see eq.~(\ref{f1})).

For the gauge field functions we obtain the equations
\begin{equation}
({\cal A}{\cal N}u_j^{'})^{'} 
 + \frac{1}{2x^2} {\cal A} (f_{j+1} - f_j) u_j = 0
\ , \label{sunu} \end{equation}
where the metric function ${\cal A}$ can be eliminated by means of
eq.~(\ref{suna}) to yield
\begin{equation}
 x^2 {\cal N} u_j^{''} + 2(\mu - x{\cal P}) u_j^{'} 
 + \frac{1}{2} (f_{j+1} - f_j) u_j = 0
\ . \label{sunu2} \end{equation}
We note the symmetry of the equations with respect to
the transformation
\begin{equation}
u_1 \rightarrow u_3  \ , \ \ \
u_3 \rightarrow u_1 
\ . \label{trans} \end{equation}

\subsection{\label{bc} Boundary conditions}

We now consider the boundary conditions
for the SU(4) EYM solutions.
Globally regular solutions must satisfy boundary conditions
at the origin. These are \cite{kuenzle}
\begin{equation}
\mu(0)=0
\ , \label{bc2} \end{equation}
and
\begin{equation}
u_j(0)=\pm 1 \ , \ \ \ j=1,2,3
\ . \label{bc3} \end{equation}
For black hole solutions with a regular horizon 
with radius $x_{\rm H}$, boundary conditions
are imposed at $x_{\rm H}$. These read
\begin{equation}
{\cal N}(x_{\rm H})=0  \ \ \ {\rm or }\ \ \ 
 \mu(x_{\rm H})= \frac{x_{\rm H}}{2}
\ , \label{bc4} \end{equation}
and
\begin{equation}
 \left. {\cal N}^{'}u_j^{'} + \frac{1}{2x^2}   (f_{j+1} - f_j) u_j 
\right|_{x_{\rm H}} =0
\ . \label{bc5} \end{equation}
For charged black hole solutions
the coefficient of $u_j^{'}$ in eq.~(\ref{bc5})
may vanish, i.~e.~${\cal N}^{'}=0$.
Then extremal black hole solutions are obtained,
which satisfy
\begin{equation}
\left. (f_{j+1} - f_{j}) u_j \right|_{x_{\rm H}} =0
\ . \label{bc6} \end{equation}

At infinity black hole solutions and globally regular solutions
satisfy the same set of boundary conditions.
Asymptotic flatness implies
that the metric functions ${\cal A}$ and $\mu$ both
approach a constant at infinity.
We here adopt
\begin{equation}
{\cal A}(\infty)=1
\ , \end{equation}
thus fixing the time coordinate.
The mass of the solutions is given by $\mu(\infty)$.
Magnetically neutral solutions are obtained, when
all gauge field functions are non-trivial and
satisfy the boundary conditions
\begin{equation}
u_j(\infty)=\pm 1 \ , \ \ \ j=1,2,3 \ ,
\   \label{bc0} \end{equation}
yielding an asymptotically vanishing
field strength tensor component $F_{\theta\phi}=0$.
No globally regular magnetically charged solutions exist.
Magnetically charged black hole solutions are obtained,
when one or more gauge field functions
are identically zero, $u_i \equiv 0$.
Their non-vanishing gauge field functions $u_j \ne u_i$
then approach constants $c_j \ne \pm 1$ at infinity \cite{kks5},
\begin{equation}
u_j(\infty)=  c_j 
\ , \label{bc1} \end{equation}
determined in section III.A. 

\section{Magnetically charged SU(4) EYM Black Hole Solutions}

When one or more gauge field functions are
identically zero, $u_j(x) \equiv 0$,
magnetically charged SU(4) EYM black hole solutions are obtained.
These solutions are important 
in classifying the neutral SU(4) EYM solutions in sequences
and in constructing their limiting solutions.
Therefore we here consider these charged SU(4) EYM 
black hole solutions first.
A general discussion of the magnetically charged SU(N) EYM solutions
was given previously \cite{kks5}.

\subsection{Classification of the 
magnetically charged SU(4) EYM black hole solutions}

We now classify the magnetically charged SU(4) EYM black hole solutions,
obtained within the ansatz (\ref{amu}) \cite{kks5}.
When one gauge field function 
is identically zero, $\omega_{k} \equiv 0$,
the ansatz reduces to
\begin{equation}
\begin{array}{ccc}
A_\mu^{(4)} dx^\mu & = & 
\left( 
\begin{array}{cc}
{\rm \fbox{$ A_\mu^{(k)} dx^\mu  $}} &                      \\
                     &  {\rm \fbox{$A_\mu^{(4-k)} dx^\mu $}}\\
\end{array}
\right)
 + {\cal H}_{k}
\end{array} \vspace{1.cm} 
\ , \label{amu1} \end{equation}
with  
${\cal H}_{k}  =  \frac{\cos\theta d\phi}{2 e} h_{k}$ and
\begin{equation}
\begin{array}{ccc}
{h}_{k} & = & 
\left( 
\begin{array}{cc}
{\rm \fbox{$(4-k){\bf 1}_{(k)} $}}&     \\
      &{\rm \fbox{$ -k {\bf 1}_{(N-k)}$}}   \\            
\end{array}
\right)\\
\end{array}\vspace{1.cm} 
\ . \label{h1} \end{equation}
Here $A_\mu^{(\bar N)}$ ($\bar N=k, 4-k$)
denotes the non-abelian static
spherically symmetric ansatz for the $su(\bar N)$ subalgebra of $su(4)$
(based on the $\bar N$-dimensional embedding of $su(2)$)
and ${\cal H}_{k}$ represents the ansatz for the element 
${h}_{k}$ of the Cartan subalgebra of $su(4)$.

The gauge field functions of the
$su(\bar N)$ parts of the solutions satisfy the boundary conditions
(eq.~(\ref{bc0}))
\begin{equation}
\bar u_i(\infty) = \pm 1
\ , \label{su2l} \end{equation}
corresponding to neutral $su(\bar N)$ solutions \cite{kks5}.
Identifying the non-vanishing functions $\omega_j$ of the $su(4)$
ansatz with the corresponding functions $\bar \omega_i$ of the
non-abelian $su(\bar N)$ ans\"atze,
\begin{equation}
\gamma_i u_i^2 = \bar \gamma_j \bar u_j^2
\ , \label{scale1} \end{equation}
then yields the asymptotic boundary conditions 
for the functions $u_j$ \cite{kks5},
\begin{equation}
 u_j(\infty) = c_j = \pm \sqrt{\bar \gamma_i/ \gamma_j} 
\ . \label{su2m} \end{equation}

The charge of the solutions is carried by the Cartan subalgebra part
of the gauge field. A solution based on the element $h_k$ of the
Cartan subalgebra carries magnetic charge of norm $P$,
\begin{equation}
P^2 = \frac{1}{2}{\rm Tr} \ {h}_{k}^2
\ . \label{Pa} \end{equation}
Expanding the element ${h}_{k}$ in terms of the
basis $\{\lambda_{n^2-1} \  | \  n=2,3,4 \}$, 
the charge can also be directly read off
the expansion coefficients,
\begin{equation}
{h}_{k} = \sum_{n=2}^N d_{k}^n P_{n^2-1} \lambda_{n^2-1} 
\ , \label{Pb} \end{equation}
where
$P_{n^2-1}=\sqrt{\frac{n(n-1)}{2}}$.
The expansion coefficients, corresponding to cases 2a-c,
are shown in Table~1.

By applying these considerations again to the subalgebras
$su(\bar N)$ of eq.~(\ref{amu1}), we obtain cases 1a-c of Table~1,
where two gauge field functions are identically zero.
In the special case (case 0 of Table~1) where
all gauge field functions are identically zero,
an embedded RN solution is obtained with charge of norm $P=\sqrt{10}$.

RN black hole solutions exist only for horizon radius
$x_{\rm H}\ge P$, and
the extremal RN solution has $x_{\rm H}=P$.
As first observed for SU(3) EYM theory \cite{gv},
the same is true for charged non-abelian black hole solutions.
Non-abelian black hole solutions with charge of norm $P$ exist only for 
horizon radii $x_{\rm H} \ge P$.
For extremal black hole solutions ${\cal N}^{'}=0$,
so the coefficient of $u_j^{'}$ in eq.~(\ref{bc5})
vanishes. This yields the boundary conditions \cite{kks5}
\begin{equation}
u_j(x_{\rm H})= c_j
\ , \label{bc8} \end{equation}
corresponding to
$\bar u_i(x_{\rm H})= \pm 1 $.

\subsection{Numerical solutions}

All possible cases of magnetically charged
SU(4) EYM black hole solutions are classified in Table~1.
Since several cases are equivalent,
1a $\sim$ 1b $\sim$ 1c and 2a $\sim$ 2c,
three non-equivalent non-abelian cases remain, 
which are discussed below.

The magnetically charged non-abelian static spherically symmetric 
solutions of SU(4) EYM theory are labelled
by the nodes of their non-vanishing gauge field functions,
i.~e.~by the nodes of the subsets of gauge field functions
$\bar u_i$, belonging to the subalgebras $su(2)$ or $su(3)$.
These solutions form sequences, completely analogously
to the sequences of neutral solutions in SU(2) and SU(3) EYM theory,
which are classified in sequences by their node structure, 
$n$ for SU(2) and $(j,j+n)$ and $(n,n)$ for SU(3).
With increasing $n$, the sequences of neutral solutions 
converge to charged solutions.
Similarly
with increasing $n$, the sequences of charged solutions 
converge to limiting solutions with higher charge.
When the limiting solution has charge of norm $P$,
then one of the gauge field functions $\bar u_i$ becomes
identically zero for $x>P$ in the limiting solution.
In particular,
the limiting solutions of case 1a are embedded RN solutions 
with charge of norm $P=\sqrt{10}$ for $x>P$.
In contrast, 
the limiting solutions of cases 2a and 2b are non-abelian solutions,
which carry charge of norm $P=3$ and
whose $su(2)$ gauge field function has $j$ nodes.
Here only for $j=0$ embedded RN solutions
with charge of norm $P=3$ are obtained for $x>P$.

In Table~2 we show the mass $\mu(\infty)$
of the first few solutions of several sequences 
together with the mass of their limiting solutions.
For every case of Table~1, two sequences are shown:
one sequence with extremal horizon
corresponding to $x_{\rm H}=\sqrt{9}$,
$x_{\rm H}=\sqrt{6}$ and $x_{\rm H}=\sqrt{8}$ for cases
1a, 2a and 2b, respectively,
and one sequence with horizon radius $x_{\rm H}=\sqrt{10}$.
For cases 2a and 2b, the second non-vanishing function
of the sequences shown has no node.
Therefore, in all three cases, the limiting solutions are embedded
RN solutions for $x>P$, where $P=\sqrt{10}$ for case 1a, and
$P=3$ for cases 2a and 2b.

We demonstrate the convergence of the functions 
for the extremal solutions of case 1a in Figs.~1.
Fig.~1a shows the gauge field function $\bar u_{1}$
for $n=1-7$ with odd $n$.
Since the limiting solution has charge of norm $P=\sqrt{10}$,
we need to distinguish two regions,
an inner region $3<x<\sqrt{10}$, beginning at the horizon,
and an outer region $x>\sqrt{10}$.
Satisfying the boundary condition
$\bar u_{1}(x_{\rm H})=1$, the function $\bar u_{1}$ becomes 
increasingly steep with increasing $n$ in the inner region,
reaching a limiting solution
whose first node resides just before $x = \sqrt{10}$,
the norm of the charge of the limiting solution.
Beyond $x = \sqrt{10}$ the function approaches zero
in an exponentially increasing region.
In Fig.~1b we show the metric function $\cal N$.
With increasing node number $n$ the function
$\cal N$ approaches a second zero, located at $x=\sqrt{10}$,
the charge of the limiting solution.
Beyond $x=\sqrt{10}$ the function $\cal N$ approaches the metric
function of the extremal RN solution with charge $P=\sqrt{10}$.
In Fig.~1c we show the charge function $P(x)$,
\begin{equation}
P^2(x) = 2x \left( \mu(\infty) - \mu(x) \right)
\ . \label{pquad} \end{equation}
With increasing $n$ the charge function $P(x)$ tends
to the charge of the limiting solution in an exponentially increasing
region.

All this is strongly reminiscent of the behaviour of the
neutral globally regular SU(2) EYM solutions and their limiting
solution \cite{bfm,kks4}.
The limiting solution of the neutral globally regular solutions 
has charge of norm $P=1$.
It also consists of two regions,
an inner region $0<x<1$, beginning at the origin,
and an outer region $x>1$.
Satisfying the boundary condition
$u(0)=1$, the function $u$
becomes increasingly steep with increasing $n$
in the inner region, reaching a limiting solution
whose first node resides just before $x = 1$,
the charge of the limiting solution.
Beyond $x=1$, the function $u$ approaches zero
in an exponentially increasing region.
Similarly, with increasing node number $n$, the function
$\cal N$ appoaches a zero located at $x=1$,
the charge of the limiting solution,
whereas beyond $x=1$ it approaches
the metric function of the extremal RN solution with charge $P=1$.

 From the above similarity between the neutral globally regular
solutions and their limiting solution
with the charged extremal solutions and their limiting solution,
we conjecture, that the charged extremal black hole solutions 
here play the role of the neutral globally regular solutions
in determining the convergence properties of the functions.
Considering neutral black hole solutions, we observed previously,
that the location of
the innermost node of the corresponding globally regular solutions
with respect to the black hole event horizon
is strongly indicative about the degree of convergence
of the black hole solutions
to their corresponding limiting solutions \cite{kks4}.

As for the globally regular solutions, the convergence of
the location of the innermost node of the charged extremal black hole
solutions is exponential.
This is seen in Fig.~2, where we present $\Delta z^1_n$ as a function
of the node number $n$, with $\Delta z^1_n$ being
defined as the deviation of the location of the
innermost node of the $n$-th extremal black hole solution 
from the location of the innermost node of the limiting solution,
\begin{equation}
\Delta z^1_n = |z^1_n - z^1_\infty|
\ . \label{Delta} \end{equation}
Considering the logarithm of $\Delta z^1_n$,
we observe, that its slope is the same for the solutions of 
case 1a and case 2b.
In contrast, the slope is different for the solutions of case 2a.
Interestingly, for the globally regular neutral SU(2) EYM solutions
the logarithm of $\Delta z^1_n$ 
also has the same slope as for the solutions of cases 1a and 2b.
We conclude, that it is only the (sub)algebra of the non-abelian
solutions, which determines the slope.

Similarly, the deviation of the mass of the $n$-th extremal 
black hole solution from the mass of the corresponding limiting solution,
\begin{equation}
\Delta \mu_n = \mu_\infty(\infty) - \mu_n(\infty)
\ , \label{Dmass} \end{equation}
decays with the same exponent, 
when the extremal solutions correspond to the same
non-abelian (sub)algebra, as seen in Fig.~3,
where in addition to cases 1a and 2b 
also the extremal charged SU(5) EYM solutions
(with $su(2)$ subalgebra) and
the globally regular SU(2) EYM solutions are shown.

\section{Regular SU(4) Einstein-Yang-Mills Solutions}

Here we consider the static spherically symmetric 
globally regular solutions of SU(4) EYM theory, 
based on the ansatz (\ref{amu}).
These discrete \cite{foot2} globally regular 
SU(4) EYM solutions are magnetically neutral.
They are analogous to the globally regular
SU(2) and SU(3) EYM solutions, obtained previously \cite{bm,kks4}.
The SU(4) EYM solutions
can be labelled by the node numbers $(n_1,n_2,n_3)$ of the
gauge field functions $u_1$, $u_2$ and $u_3$. The solutions
can then be classified into sequences.
We first discuss several sequences which have embedded
abelian limiting solutions with magnetic charge of norm $P$ for $x>P$.
Then we discuss the general case and classify the general solutions
into sequences, presenting some numerical examples.

\subsection{Sequences with embedded abelian limiting solutions}

Here we discuss those sequences of globally regular solutions,
which for $x>P$ tend to embedded abelian limiting solutions
with magnetic charge of norm $P$.
In these sequences a single index $n$
characterizes the number of nodes of one, two or all three
gauge field functions, while the remaining gauge field functions
have zero nodes.
With increasing node number $n$ the gauge field functions 
with zero nodes tend towards some finite constant value for $x>P$,
whereas the gauge field functions 
with $n$ nodes tend to limiting functions, which vanish for $x>P$.
Consequently the limiting solutions represent
embedded charged abelian black holes for $x>P$, whose charge may
be read off Table~1.

\subsubsection{$(n,0,0)$ sequence}

In Figs.~4a-c we present the first few globally regular solutions 
of the sequence with node structure $(n,0,0)$, $n=1-7$, with odd $n$.
The first member of the $(1,0,0)$ sequence is the solution with
the lowest mass of all SU(4) EYM solutions
(within the ansatz (\ref{amu})), $\mu(\infty)=1.723995$ (see Table~3). 

The limiting solution of this sequence is classified by case 2c of Table~1.
For $n \rightarrow \infty$, the limiting solutions
corresponding to case 2c of Table~1
are EYM solutions based on the $su(3)$ subalgebra of $su(4)$,
which carry magnetic charge of norm $P=\sqrt{6}$.
For $x>P$ the gauge field function $u_1$ 
of the solutions of this sequence tends to zero, while
the functions $u_2$ and $u_3$ approach
the SU(3) vacuum solution. Therefore for $x>P$ 
in this case the limiting solution
corresponds to an embedded abelian solution,
an extremal RN solution with the same charge (of norm $P=\sqrt{6}$).

In Fig.~4a the gauge functions $u_1$, $u_2$ and $u_3$ are presented.
For $x>P$ the limiting gauge field
functions are $u_1 \equiv 0$, 
$u_2\equiv \sqrt{1/2}$ and $u_3\equiv \sqrt{2/3}$,
whereas for $x<P$ they are non-trivial functions.
With increasing node number $n$,
the location of the innermost node of the function $u_1$ 
tends exponentially to a finite limiting value and
an accumulation point appears at $x= P$.

In Fig.~4b we present the metric function $\cal N$. 
For small node number $n$ ($n\le 3$) $\cal N$ possesses only one minimum, 
while for larger $n$ several local minima of $\cal N$ develop. 
For $n \rightarrow \infty$,
the global minimum ${\cal N}|_{\rm min}=0$ occurs
at $x({\cal N}|_{\rm min})= P$.
It represents the event horizon of the embedded extremal RN solution,
which corresponds to the limiting solution for $x>P$.

In Fig.~4c we present the charge function $P(x)$. 
For finite node number $n$ all solutions
are neutral, and $P(x)$ decays asymptotically to zero.
However, with increasing $n$ the charge function $P(x)$ tends to
the charge of the limiting solution $P=\sqrt{6}$
in an exponentially increasing region.

Since the equations of motion are symmetric under the interchange 
of $u_1$ and $u_3$, the solutions with node structure $(0,0,n)$ 
form an equivalent degenerate sequence.

\boldmath
\subsubsection{Other sequences}
\unboldmath

We now consider the other sequences 
of globally regular solutions which
tend to embedded abelian solutions with charge of norm $P$ for $x>P$.
Turning to the sequence with node structure $(0,n,0)$,
the limiting solution of this sequence is classified by case 2b of Table~1.
The limiting solutions corresponding to case 2b of Table~1
carry magnetic charge of norm $P=\sqrt{8}$. For $x>P$ they represent
EYM solutions based on the $su(2) \oplus su(2)$ subalgebra of $su(4)$.
Since the gauge field functions 
$u_1$ and $u_3$ of the solutions of this sequence approach
the SU(2) vacuum solution, in this case the limiting solution
corresponds to an embedded abelian solution for $x>P$,
an extremal RN solution with charge $P=\sqrt{8}$.
The masses $\mu_n(\infty)$ of this sequence are shown in Table~3.
In addition to this sequence with $u_1 = u_3$, 
a second sequence with node structure $(0,n,0)$ might exist with
$u_1 \ne u_3$.
However, we do not find such solutions numerically.

Next we consider the cases where two gauge field functions have a finite
number of nodes and the third function has zero nodes. 
The first such case is the sequence with node structure $(n,n,0)$.
The limiting solution of this sequence is classified by case 1c of Table~1.
The limiting solutions corresponding to case 1c of Table~1
carry magnetic charge of norm $P=\sqrt{9}$. For $x>P$ they represent
EYM solutions based on the $su(2)$ subalgebra of $su(4)$.
Since the gauge field function $u_3$ of the 
solutions of the $(n,n,0)$ sequence approaches
the SU(2) vacuum solution, in this case the limiting solution
corresponds to 
an embedded extremal RN solution with charge $P=\sqrt{9}$ for $x>P$.
The masses $\mu_n(\infty)$ of this sequence are shown in Table~3
\cite{foot3}.
An equivalent degenerate sequence is $(0,n,n)$.

Considering solutions with node structure $(n,0,n)$,
we find two types of solutions 
leading to two distinct sequences. In the solutions of the first type
the gauge functions $u_1$ and $u_3$ are identical,
whereas in the second type $u_1 \ne u_3$.
The limiting solutions of both sequences are identical and
classified by case 1b of Table~1.
Analogously to case 1c,
the limiting solutions corresponding to case 1b of Table~1
carry magnetic charge of norm $P=\sqrt{9}$
and represent EYM solutions for $x>P$,
based on the $su(2)$ subalgebra of $su(4)$.
Only for the $(n,0,n)$ sequences the limiting solution
corresponds to an embedded extremal RN solution with $P=\sqrt{9}$ for $x>P$.
The masses $\mu_n(\infty)$ of both of these sequences
are shown in Table~3.

Last we consider the solutions with node structure $(n,n,n)$.
Here three distinct types of solutions exist.
In the solutions of the first type
all three gauge field functions are identical,
$u_1=u_2=u_3$. These simply represent scaled SU(2) solutions,
with scaling factor $\sqrt{10}$.
In the solutions of the second type
only two gauge field functions are identical, $u_1 = u_3$,
and in the solutions of the third type all three gauge field functions
differ from each other.
For the second and third type of solutions, for a given $n$ 
several distinct non-degenerate solutions may exist.
These solutions are discussed in detail in section~V.B.
The limiting solutions of the $(n,n,n)$ sequences are
classified by case 0 of Table~1.
Thus for $n \rightarrow \infty$, the sequences tend to an embedded
extremal RN solution with magnetic charge of norm $P=\sqrt{10}$ for $x>P$.
The masses $\mu_n(\infty)$ for several $(n,n,n)$ sequences
are shown in Table~4.

\boldmath
\subsection{General sequences}
\unboldmath

All sequences of neutral globally regular solutions
not included in the previous subsection 
tend to non-abelian limiting solutions for $x>P$,
which carry magnetic charge of norm $P$.
The norm of the charge of the limiting solutions
is the same as in the corresponding abelian cases.
In this subsection we
first present a classification scheme for the general 
globally regular solutions of 
SU(4) EYM theory (based on the ansatz (\ref{amu}))
and then demonstrate the main features of these solutions
by constructing several sequences numerically.

\subsubsection{Classification of the general solutions}

The general globally regular solutions are labelled by the node numbers
$(n_1,n_2,n_3)$ of the corresponding gauge field functions $u_1,u_2,u_3$.
The case, where all node numbers are identical,
has already been considered in the previous subsection. 
Assuming next, that the node numbers of any two gauge field functions 
($u_j$) are identical, we obtain four types of sequences, 
labelled by the fixed index $k$ and the running index $n$.
For fixed $k$, $k=1$, 2, 3,..., and running $n$, $n=1$, 2, 3,...,
these four types of sequences are
$(k,k,n+k)$, $(k,n+k,k)$, $(k,n+k,n+k)$ 
and $(n+k,k,n+k)$.
For $k=0$ the sequences of the previous subsection are obtained.
In this scheme any solution with two identical node numbers
is included exactly once, apart from equivalent degenerate solutions
and the cases, where solutions with both $u_1 \ne u_3$ and 
$u_1 = u_3$ exist.
The limiting solutions of the sequences
$(k,n+k,n+k)$ and $(n+k,k,n+k)$ correspond to cases 1a and 1b, respectively, 
and the limiting solutions of the sequences
$(k,n+k,k)$ and $(k,k,n+k)$ correspond to cases 2b and 2a, respectively 
(Table~1).

When all three node numbers differ from each other, we obtain a similar
classification, which now requires three distinct indices,
$l$, $k$ and $n$. 
For fixed indices $l$ and $k$ with $l=0$, 1, 2,...,
and $k=1$, 2, 3,...,
and running index $n$,
there are three types of sequences,
$(l,l+k,l+k+n)$, $(l+k,l,l+k+n)$ and $(l,l+k+n,l+k)$.  
Any solution with three different node numbers is thus
included exactly once in this scheme, apart from
equivalent degenerate solutions.
The limiting solutions of the
$(l,l+k,l+k+n)$ and $(l+k,l,l+k+n)$ sequences
correspond to case 2a of Table~1
and the limiting solutions of the 
$(l,l+k+n,l+k)$ sequence corresponds to case 2b.

It is straightforward to extend this classification to the globally
regular solutions of SU(N) EYM theory.

\subsubsection{$(n+1,1,0)$ sequence and $(n+1,0,1)$ sequence}

We now discuss the properties of these general sequences. Their
qualitative features are analogous to those of the 
sequences discussed in the previous subsection.
In order to demonstrate this, we here present two numerically constructed
sequences: $(n+1,1,0)$ and $(n+1,0,1)$. 

In Fig.~5a we present the gauge field functions 
of the globally regular solutions of the sequence with
node structure $(n+1,1,0)$ for $n=1-5$.
The limiting solution of this sequence is an EYM solution
based on the $su(3)$ subalgebra of $su(4)$,
with magnetic charge of norm $P=\sqrt{6}$,
corresponding to case 2c of Table~1.
For $x>P$ the limiting solution is the extremal SU(3) EYM
black hole solution with node structure $(1,0)$.
In Fig.~5b the metric function ${\cal N}$ is shown.
With increasing node number $n$
the metric function ${\cal N}$ tends to zero at $x=\sqrt{6}$,
the charge of the limiting solution.
The metric function ${\cal N}$ has the same qualitative behaviour 
for the $(n+1,1,0)$ sequence as for the $(n,0,0)$ sequence.
The charge function $P(x)$ is shown in Fig.~5c. 
Again, with increasing $n$ it approaches $P=\sqrt{6}$ in an
exponentially increasing region.

The mass $\mu(\infty)$ of the solutions of the $(n+1,1,0)$ sequence
again converges exponentially to the mass of the limiting solution.
This is seen in Fig.~6, where $\Delta \mu_n$
(eq.~(\ref{Dmass})) is presented.
Also shown in Fig.~6 
is the mass of the solutions of the $(n+1,0,1)$ sequence.
For a given $n$,
the solutions of the $(n+1,0,1)$ sequence have a lower mass
than the solutions of the $(n+1,1,0)$ sequence,
but both sequences tend to the same limiting solution.
For comparison, also the mass of
the solutions of the $(n,0,0)$ sequence is shown.
All three sequences are based on the $su(3)$ subalgebra of $su(4)$,
and the function $\ln \Delta \mu_n$ has the same slope for 
all three sequences. This is analogous to the case
of the solutions based on the $su(2)$ subalgebra of $su(4)$,
discussed in section III.B.
The masses of the solutions of the $(n+1,1,0)$ and
$(n+1,0,1)$ sequences and of their limiting solution 
are shown in Table~3. 

\section{Neutral SU(4) Einstein-Yang-Mills Black Holes}

The SU(4) EYM black hole solutions are obtained
analogously to the globally regular solutions,
but with boundary conditions imposed at the event horizon
(see section \ref{bc}).
As for the SU(3) EYM black hole solutions \cite{kks2}, 
in general sequences of neutral SU(4) EYM black hole solutions 
exist for all values of the horizon radius.
However, for the SU(3) EYM solutions with node structure $(n,n)$
there are two distinct types of solutions,
the scaled SU(2) solutions with $u_1=u_2$ 
and the genuine SU(3) solutions with $u_1 \ne u_2$.
The genuine SU(3) solutions
exist only up to a critical value of the horizon
radius $x_{\rm H \, n}^{\rm cr}$, where they merge into the 
scaled SU(2) solutions.
Analogously, in SU(4) EYM theory critical values of the horizon
radius occur for the various types of
solutions with node structure $(n,j,n)$ and $(n,n,n)$.
Their complex critical behaviour is analyzed below \cite{foot3}.

\subsection{General sequences}

To all of the globally regular solutions 
discussed above, the corresponding neutral
black hole solutions exist at least for sufficiently small values of
the horizon radius $x_{\rm H}$. 
For fixed horizon radius $x_{\rm H}$,
the sequences of neutral black hole solutions 
tend to limiting solutions with charge of norm $P$.
For $x_{\rm H}<P$ the sequences of 
black hole solutions tend to a non-trivial limiting solution 
for $x_{\rm H}<x<P$ and to 
a charged non-abelian black hole solution or 
an embedded extremal RN solution for $x > P$. 
For $x_{\rm H}>P$ the sequences of 
black hole solutions tend to 
a charged non-abelian black hole solution or 
an embedded RN solution
with charge $P$ and the same value of the horizon radius.
Since the other qualitative features of the black hole solutions 
are completely analogous to those of the
corresponding globally regular solutions, 
we do not discuss them here further.
Instead we turn to the solutions with node structure
$(n,j,n)$ and $(n,n,n)$, whose complex critical behaviour
shows interesting novel features.

\subsection{Bifurcations}

We recall that
for the globally regular solutions with node structure $(n,0,n)$, 
two types of solutions exist, one type with $u_1 = u_3$ 
and a second type with $u_1 \ne u_3$, whereas
for the globally regular solutions with node structure $(n,n,n)$, 
three types of solutions exist, one type
with $u_1 = u_2 = u_3$, representing scaled SU(2) solutions,
a second type with $u_1 = u_3 \ne u_2$, and a third type
with $u_1 \ne u_3 \ne u_2$.

Similarly, for the black hole solutions with node structure $(n,0,n)$,
there are two distinct types
of black hole solutions for sufficiently small values of the event horizon.
At a critical value of the horizon radius $x_{\rm H \, n}^{\rm cr}$,
the $u_1 \ne u_3$ solutions with node number $n$ merge 
into the $u_1 = u_3$ solutions.
Beyond this critical value only the $u_1=u_3$ solutions persist.
With increasing node number $n$
the critical value of the horizon radius 
$x_{\rm H \, n}^{\rm cr}$ decreases.
For the black hole solutions with $n=1$, 3 and 5
the critical values of the horizon radius are
$x_{\rm H \, 1}^{\rm cr} = 1.972$,
$x_{\rm H \, 3}^{\rm cr} = 1.244$ and
$x_{\rm H \, 5}^{\rm cr} = 1.180$ (see Table~5).

In Fig.~7a the value of the gauge field functions at the horizon
of the black hole solutions with node structure $(n,0,n)$
is shown for $n=1$, 3 and 5, together with the corresponding
three critical values of the horizon radius.
For $n=3$ and 5 the $u_1=u_3$ solutions are only shown
in the inset, in order not to cover 
part of the corresponding $u_1 \ne u_3$ solutions.
For the $u_1 \ne u_3$ solutions, we obtain two degenerate branches
of solutions. They are degenerate
because of the symmetry with respect to the interchange of
$u_1$ and $u_3$,
i.~e.~if in a given solution $u_1(x_{\rm H})$ lies on the upper branch 
and $u_3(x_{\rm H})$ lies on the lower branch,
then in the degenerate solution $u_1(x_{\rm H})$ and $u_3(x_{\rm H})$
are interchanged.
In Fig.~7b we present the corresponding masses.
The mass of the $u_1 \ne u_3$ solutions
is always larger than the mass of the $u_1 = u_3$ solutions.
In contrast, the temperature is lower for the $u_1 \ne u_3$ solutions,
as seen in Fig.~7c.
 
We now consider the black hole solutions with node structure $(n,n,n)$,
whose critical behaviour is more complex.
As for the globally regular solutions,
three distinct types of black hole solutions exist
for sufficiently small values of the horizon radius.
There is one solution of the first type with $u_1 = u_3 = u_2$,
we find two non-degenerate solutions of the second type with
$u_1 = u_3 \ne u_2$, 
and we find up to four non-degenerate solutions of the third type with
$u_1 \ne u_3 \ne u_2$.
At the largest critical value of the horizon radius 
$x_{\rm H \, n}^{\rm cr,1}$,
the two non-degenerate branches of 
$u_1 = u_3 \ne u_2$ solutions merge into each other. Notably, 
this critical value of the horizon radius $x_{\rm H \, 1}^{\rm cr,1}$
is slightly beyond the value of the horizon radius,
where the branches intersect the branch of scaled SU(2) solutions.
Beyond $x_{\rm H \, n}^{\rm cr,1}$ only the scaled SU(2) solutions
persist.
At the second largest critical value of the horizon radius
$x_{\rm H \, n}^{\rm cr,2}$, two degenerate branches of
$u_1 \ne u_3 \ne u_2$ solutions
merge into the branch of scaled SU(2) solutions. 
At the third critical value
$x_{\rm H \, n}^{\rm cr,3}$ another two degenerate branches of 
$u_1 \ne u_3 \ne u_2$ solutions
merge into the lower branch of the $u_1 = u_3 \ne u_2$ solutions. 
At the smallest fourth critical value
$x_{\rm H \, n}^{\rm cr,4}$ two non-degenerate branches of 
$u_1 \ne u_3 \ne u_2$ solutions
merge into each other,
being neither close to the branch of scaled SU(2) solutions or
one of the $u_1 = u_3 \ne u_2$ branches.
They cease to exist beyond $x_{\rm H \, n}^{\rm cr,4}$.

We illustrate this complex critical behaviour for
the black hole solutions with node structure (1,1,1) in Figs.~8.
The value of the gauge field functions 
at the horizon radius $x_{\rm H}$ is shown in Fig.~8a
for all solutions found,
together with the four critical values of the horizon radius.
The scaled SU(2) solutions, having $u_1=u_3=u_2$,
exist for all values of the horizon radius.
We see two distinct non-degenerate branches of 
$u_1 = u_3 \ne u_2$ solutions, 
up to the critical value of the
horizon radius $x_{\rm H \, 1}^{\rm cr,1} = 4.158$.
Of the six branches of $u_1 \ne u_3 \ne u_2$ solutions
two degenerate branches 
of solutions merge into the branch of scaled SU(2) solutions
at the critical value $x_{\rm H \, 1}^{\rm cr,2} = 2.714$,
another two degenerate branches 
merge into the lower branch of the $u_1 = u_3 \ne u_2$ solutions
at the critical value $x_{\rm H \, 1}^{\rm cr,3} = 1.025$, 
as seen in the upper right inset of Fig.~8a,
and two non-degenerate branches merge into each other
at $x_{\rm H \, 1}^{\rm cr,4} = 0.445$,
as seen in the lower left inset of Fig.~8a.
In Fig.~8b we present the mass of all
branches of solutions as a function of the horizon radius.
Of all seven distinct branches of solutions, the scaled SU(2) 
solutions have the lowest mass, and first one and then the other
of the two branches of $u_1 = u_3  \ne u_2$ solutions
has the highest mass, except very close to 
$x_{\rm H \, 1}^{\rm cr,1}$.
The intersection of the two $u_1 = u_3 \ne u_2$ branches
is seen clearly in Fig.~8c, where the temperature 
of all branches is shown.

The critical behaviour of the black hole solutions 
with node structure $(3,3,3)$ is shown in Fig.~9.
The two non-degenerate branches of $u_1 = u_3 \ne u_2$ solutions
cease to exist at the critical value
$x_{\rm H \, 3}^{\rm cr,1} = 2.836$,
as seen in the upper right inset,
two degenerate branches 
of $u_1 \ne u_3 \ne u_2$ solutions merge into the branch of
scaled SU(2) solutions
at the critical value $x_{\rm H \, 3}^{\rm cr,2} = 2.090$,
as seen in the lower left inset,
and two degenerate branches 
of $u_1 \ne u_3 \ne u_2$ solutions merge into the lower branch of
$u_1 = u_3 \ne u_2$ solutions at $x_{\rm H \, 3}^{\rm cr,3} = 0.474$.
The critical values of the horizon radius and
the corresponding masses of the $(n,n,n)$ sequences
are presented in Table~4 for $n=1-5$ \cite{foot4}.

\section{SU(4) Einstein-Yang-Mills-Dilaton Solutions}

We here briefly consider the solutions of SU(4) EYMD theory. 
In EYMD theory the dilaton coupling constant $\gamma$
represents an additional parameter.
For $\gamma=0$ EYM theory is recovered,
whereas for $\gamma=1$ contact with the low-energy effective action of 
string theory is made.
Most of the qualitative features
of the static spherically symmetric EYMD solutions agree with those
of the static spherically symmetric EYM solutions.
This was observed previously for the solutions of 
SU(2) and SU(3) EYMD theory \cite{eymd,gal,kks3,kks4},
and it also holds for the solutions of SU(4) EYMD theory.
In particular, the solutions of SU(4) EYMD theory can 
be classified in the same way as the SU(4) EYM solutions.
Qualitative differences concern the charged black hole solutions
and the limiting solutions of the sequences of solutions.
As in EMD theory, where only the extremal solution
with $x_{\rm H}=0$ has a naked singularity at the origin,
in EYMD theory charged black hole solutions 
with a regular horizon exist for any
value of the horizon radius $x_{\rm H}>0$.
The limiting solutions of the sequences of EYMD black hole solutions
are EYMD and embedded 
EMD black hole solutions with the same horizon radius and the
same dilaton coupling constant, 
which carry the same charge $P$ as the limiting solutions of
the corresponding sequences of EYM solutions,
as demonstrated in detail for SU(3) EYMD theory \cite{kks4}.
The limiting solutions of the sequences of globally regular
EYMD solutions are extremal EYMD and 
embedded EMD solutions \cite{eymd,kks4}.

\subsection{Action and Equations}

We briefly discuss the action, the equations of motion
and the boundary conditions for the static spherically
symmetric solutions of SU(4) EYMD theory.
The SU(4) EYMD action is
\begin{equation}
S=S_G+S_M=\int L_G \sqrt{-g} d^4x + \int L_M^D \sqrt{-g} d^4x
\ , \label{action2}  \end{equation}
with $L_G$ given in eq.~(\ref{LG}) and
\begin{equation}
L_M^D=-\frac{1}{2}\partial_\mu \Phi \partial^\mu \Phi
 -\exp({2 \kappa \Phi })\frac{1}{2} {\rm Tr} (F_{\mu\nu} F^{\mu\nu})
\ , \label{lagmd} \end{equation}
with field strength tensor
$F_{\mu\nu}$ given in eq.~(\ref{FMN}), and 
dilaton coupling constant $\kappa$.

In EYMD theory we proceed analogously to EYM theory.
We employ Schwarz\-schild-like coordinates and adopt
the static spherically symmetric metric (\ref{metric}). 
The ansatz for the static spherically symmetric 
gauge field $A_\mu$ is given by (\ref{amu}). 
The corresponding ansatz for the static spherically symmetric
dilaton field is $\Phi=\Phi(r)$.
With these ans\"atze we obtain for the matter Lagrangian
\begin{eqnarray}
L_M^D &=& -\frac{1}{2} {\cal N} \Phi'^2
    - \frac{\exp({2 \kappa \Phi})}{e^2 r^2} 
    \left( {\cal N} {\cal G} + {\cal P} \right)
\ , \end{eqnarray}
with ${\cal N}$, ${\cal G}$ and ${\cal P}$ given in
(\ref{n}) and (\ref{gp}).

Changing to the dimensionless coordinate $x$ given in (\ref{x}) 
and the dimensionless mass function $\mu(x)$ given in (\ref{mu}),
we further introduce the dimensionless dilaton field $\phi$,
\begin{equation}
\phi = \sqrt{4\pi G} \Phi
\  , \end{equation}
and the dimensionless coupling constant $\gamma$,
\begin{equation}
\gamma =\kappa/\sqrt{4\pi G}
\  . \end{equation}
The choice $\gamma=1$ corresponds to string theory,
while $4+n$ dimensional Kaluza-Klein theory
has $\gamma^2=(2+n)/n$ \cite{emd}.

 From the Einstein equations we obtain the field equations 
for $\mu(x)$ and ${\cal A}(x)$,
\begin{eqnarray}
\mu^{'} & = & \frac{1}{2}{\cal N} x^2 \phi'^2 + 
 \exp({2 \gamma \phi }) \left({\cal N}\cal{G}  + \cal{P}\right)
   \ , \label{sundm}\\
\frac{{\cal A}^{'}}{{\cal A}} & = & \frac{2}{x} \left(
 \frac{1}{2} x^2 \phi'^2+ \exp({2 \gamma \phi })  {\cal G} \right)
\ , \label{eqd1}\end{eqnarray}
where the prime now indicates the derivative with respect to $x$.

For the gauge field functions $u_j(x)$, $j=1-3$,
and for the dilaton function $\phi(x)$
we obtain the field equations
\begin{eqnarray}
(\exp({2 \gamma \phi }){\cal A} {\cal N}  u_j')'
=- \frac{\exp({2 \gamma \phi })}{2 x^2}{\cal A} (f_{j+1}-f_j) u_j \ , \\
({\cal A} {\cal N} x^2 \phi')' = 2 \gamma {\cal A}
 \exp({2 \gamma \phi }) \left( {\cal N}\cal{G}  + \cal{P}\right)
\ . \label{eqd2} \end{eqnarray}

The metric and gauge field
functions satisfy the same boundary conditions as in EYM theory, 
given in section II.D.
These are supplemented with the boundary conditions for the dilaton function. 
At infinity the dilaton function satisfies
\begin{equation}
\phi(\infty) =0
\ , \end{equation}
for the globally regular solutions 
the boundary condition at the origin is
\begin{equation}
\phi'(0) =0
\ , \end{equation}
and for the black hole solutions regularity at the horizon requires
\begin{equation}
 {\cal N}' \phi' = \frac{2 \gamma \exp({2 \gamma \phi }) }{x^2}
  \left( {\cal N}\cal{G}  + \cal{P}\right)
\ . \end{equation}

\subsection{Numerical solutions}

In SU(4) EYMD theory, sequences of neutral static spherically symmetric 
globally regular solutions can be constructed 
for arbitrary dilaton coupling constant $\gamma$. 
Parameterized by the node numbers $(n_1,n_2,n_3)$ 
of the gauge field functions $(u_1,u_2,u_3)$, 
these sequences converge to charged static spherically 
symmetric extremal EYMD solutions, based on the subalgebras of $su(4)$
as classified in Table~1 for EYM theory.

We present some examples of globally regular solutions in
the upper part of Table~5.
Shown is the mass of the first few solutions (with odd $n$) 
of the sequences with node structure
$(n,0,n)$ with $u_1=u_3$ and $u_1 \ne u_3$
for EYMD theory with $\gamma=1$ and for EYM theory ($\gamma=0$).
Also shown is the mass of the corresponding limiting solutions,
representing an embedded extremal EMD solution for EYMD theory
and an embedded extremal RN solution for $x>P$ for EYM theory.

Analogously, sequences of neutral static spherically symmetric 
black hole solutions can be constructed 
for arbitrary dilaton coupling constant $\gamma$. 
These sequences converge to the corresponding
charged static spherically symmetric EYMD and 
embedded EMD black hole solutions.
The temperature shown in the upper part of Table~5 
for the globally regular solutions should be interpreted
as the temperature of the corresponding black hole solutions
in the limit $x_{\rm H} \rightarrow 0$.

The lower part of Table~5 shows the mass and temperature
of the black hole solutions of the $(n,0,n)$ sequences
at the critical values of the horizon radius $x_{\rm H \, n}^{\rm cr}$, 
where the $u_1 \ne u_3$ solutions merge into the $u_1=u_3$ solutions,
for EYMD theory with $\gamma=1$ and for EYM theory ($\gamma=0$).
Also shown are the critical values themselves 
and the value of the function $u_1$ at $x_{\rm H \, n}^{\rm cr}$.
For $\gamma=1$ the critical values are smaller than for $\gamma=0$.
This is in agreement with the critical values of the SU(3) EYMD
black hole solutions \cite{kks4}, 
which with increasing $\gamma$ first decrease and then increase again.

The charged EYMD black hole solutions also follow the classification
scheme of the corresponding charged EYM black hole solutions.
Since these were discussed at length for SU(3) EYMD theory in
\cite{kks4}, we do not present further results here.

\section{Conclusions}

We have investigated the various sequences of globally regular
and black hole solutions of SU(4) EYM theory,
based on the static spherically symmetric ansatz for the 
purely magnetic gauge field,
obtained by embedding the 4-dimensional representation of 
$su(2)$ in $su(4)$ \cite{kuenzle}.
We have classified the discrete set of
solutions in sequences by means of the node numbers
$(n_1,n_2,n_3)$ of their gauge field functions $u_1$, $u_2$
and $u_3$. 
We have determined the limiting solutions of the sequences of solutions,
and we have constructed numerous sequences and their limiting
solutions numerically.
In particular, we have shown, that the exponential convergence
of the mass of the globally regular and extremal black hole
solutions depends on the non-abelian (sub)algebra
of the solutions.

The SU(4) EYM black hole solutions with node structure
$(n,j,n)$ and $(n,n,n,)$ show a complex critical behaviour.
For such sequences several branches of black hole solutions exist
up to critical values of the horizon radius.
For sufficiently small values of the horizon radius
two types of solutions with node structure $(n,0,n)$ exist.
The $u_1 \ne u_3$ solutions merge into the $u_1 = u_3$ solutions
at a critical value of the horizon radius $x_{\rm H \, n}^{\rm cr}$.
Beyond $x_{\rm H \, n}^{\rm cr}$ only the 
$u_1 = u_3$ solutions persist.
Again for sufficiently small values of the horizon radius
there are three types of solutions with node structure $(n,n,n)$,
scaled SU(2) solutions with $u_1 = u_2 = u_3$, 
solutions with $u_1 = u_3 \ne u_2$, and solutions
with $u_1 \ne u_3 \ne u_2$.
The two non-degenerate branches of 
$u_1 = u_3 \ne u_2$ solutions merge into each other
and cease to exist beyond $x_{\rm H \, n}^{\rm cr,1}$.
At $x_{\rm H \, n}^{\rm cr,2}$, two degenerate branches of 
$u_1 \ne u_3 \ne u_2$ solutions
merge into the branch of scaled SU(2) solutions. 
Another two degenerate branches of $u_1 \ne u_3 \ne u_2$ solutions
merge into the lower branch of the $u_1 = u_3 \ne u_2$ solutions
at $x_{\rm H \, n}^{\rm cr,3}$,
and two non-degenerate branches of $u_1 \ne u_3 \ne u_2$ solutions
merge into each other at $x_{\rm H \, n}^{\rm cr,4}$ \cite{foot4}.

For SU(4) EYM theory, the classification of the solutions in sequences
represents a generalization with respect to SU(3) EYM theory, since
three node numbers are needed for the classification instead of two.
It is straightforward to generalize the classification 
of the static spherically symmetric globally regular
and black hole solutions further to SU(N) EYM theory, 
where $(N-1)$ node numbers are needed for the classification
since $(N-1)$ gauge field functions are present.
For the charged SU(N) EYM black hole solutions, 
the general classification is given in \cite{kks5}.
An existence proof for regular SU(N) EYM solutions 
has been proposed recently \cite{proofn}, which follows closely 
the existence proof for regular SU(2) EYM solutions \cite{proof2}.

For EYMD theory the same classification of the solutions holds.
In EYM theory charged black hole solutions exist only
for horizon radius $x_{\rm H} \ge P$, where $P$ is the norm of the charge,
whereas in EYMD theory charged black hole solutions 
exist for any value of the horizon radius $x_{\rm H}>0$
as in EMD theory.
The limiting solutions of the sequences of EYMD black hole solutions
are charged EYMD and embedded 
EMD black hole solutions with the same horizon radius,
whereas the limiting solutions of sequences of EYM black hole solutions
are charged EYM and embedded RN black hole solutions 
with the same horizon radius for $x_{\rm H} > P$.
The norm of the charge of the limiting solutions is determined
by their $su(4)$ subalgebras and is the same for
EYM and EYMD theory.

Recently, new types of solutions
were found in SU(2) EYM and EYMD theory.
These are static globally regular and black hole solutions
with only axial symmetry \cite{kk2,kk3}
and non-static non-rotational black hole solutions
\cite{brod,slomo}.
Such solutions as well as solutions with
only discrete symmetries \cite{ewein}
are also expected in SU(N) EYM and EYMD theory with $N>2$.

{\sl Acknowledgement}

The work of M. Wirschins was supported in part by the DFG.

\newpage
 \newcommand{\rb}[1]{\raisebox{1.5ex}[-1.5ex]{#1}}
\begin{table}[p!]
\begin{center}
\begin{tabular}{|c|ccc|ccc|c|l|p{0.75cm}p{0.75cm}p{0.75cm}|} \hline
 & & & & & & & & \multicolumn{1}{ c|}{non-abelian}&
 \multicolumn{3}{ c|}{Cartan}\\
 & & & & & & & & \multicolumn{1}{ c|}{subalgebra} & 
 \multicolumn{3}{ c|}{subalgebra $^*$} \\
\hline
\# & $u_1$ & $u_2$ & $u_3$ &
$c_1^2$       & $c_2^2$       & $c_3^2$       &
$P^2$ &  &
  $\lambda_3$ & $\lambda_8$ & $\lambda_{15}$ \\
 \hline
0  & 0     & 0     & 0     &
0             & 0             & 0             &
10 &       &
$P_{3}$  & $P_{8}$ & $P_{15}$ \\
 \hline
1a & $u_1$ & 0     & 0     &
$\frac{1}{3}$ & 0             & 0             &
 9 & $su(2)$ &
         & $P_{8}$ & $P_{15}$ \\
1b & 0     & $u_2$ & 0     &
0             & $\frac{1}{4}$ & 0             &
 9 & $su(2)$ &
$\frac{3}{2}P_{3}$ & $\frac{1}{2}P_{8}$ & $P_{15}$ \\
1c & 0     & 0     & $u_3$ &
0             & 0             & $\frac{1}{3}$ &
 9 & $su(2)$ &
$P_{3}$ & $\frac{4}{3}P_{8}$ & $\frac{2}{3}P_{15}$ \\
 \hline
2a & $u_1$ & $u_2$ & 0     &
$\frac{2}{3}$ & $\frac{1}{2}$ & 0             &
 6 & $su(3)$ &
         &         & $P_{15}$ \\
2b & $u_1$ & 0     & $u_3$ &
$\frac{1}{3}$ & 0             & $\frac{1}{3}$ &
 8 & $su(2)\oplus su(2)$ &
        & $\frac{4}{3}P_{8}$ & $\frac{2}{3}P_{15}$ \\
2c & 0     & $u_2$ & $u_3$ &
0             & $\frac{1}{2}$ & $\frac{2}{3}$ &
 6 & $su(3)$ &
$2P_{3}$ & $\frac{2}{3}P_{8}$ & $\frac{1}{3}P_{15}$ \\
\hline
\end{tabular}
\end{center} 
\vspace{1.cm} 
{\bf Table 1}\\
The classification of the charged black hole solutions of SU(4)
EYM theory is presented.
Shown are the non-vanishing gauge field functions (denoted by $u_j$)
and the identically vanishing gauge field functions (denoted by zero),
the norm squared of the charge of the black hole solutions, $P^2$,
and the subalgebra of the solutions including the non-abelian subalgebra
and the coefficients$^*$ (in the given basis) of the corresponding 
element of the Cartan subalgebra.
\end{table}
\newpage
\begin{table}[p!]
\begin{center}
\begin{tabular}{|c|ccc|} \hline
\multicolumn{1}{|c|} { $ $ }&
\multicolumn{3}{ c|} {  $\mu(\infty)$ }\\
$ $ & 1a     & 2a     & 2b         \\
\hline
$n/x_{\rm H}$& $\sqrt{9}$ & $\sqrt{6}$ & $\sqrt{8}$  \\
 \hline    
$1$      & 3.12998 & 2.83874 & 2.96587\\
$2$      & 3.15685 & 2.95799 & 2.99427 \\
$3$      & 3.16139 & 2.98951 & 2.99906 \\
$4$      & 3.16213 & 2.99741 & 2.99985\\
$5$      & 3.16225 & 2.99936 & 2.99997\\
$7$      & 3.16228 & 2.99996 & 3.00000 \\
$\infty$ & 3.16228 & 3.0  & 3.0 \\
 \hline
\multicolumn{1}{|c|} { $ $ }&
\multicolumn{3}{ c|} {  $x_{\rm H}=\sqrt{10}$ } \\
 \hline
$1$      & 3.13343 & 2.89417 & 2.98031\\
$2$      & 3.15864 & 2.98734 & 2.98734 \\
$3$      & 3.16191 & 3.00255 & 3.00412 \\
$4$      & 3.16224 & 3.00406 & 3.00416\\
$5$      & 3.16226 & 3.00415 & 3.00416\\
$7$      & 3.16228 & 3.00416 & 3.00416\\
$\infty$ & 3.16228 & 3.00416 & 3.00416\\
 \hline
\end{tabular}
\end{center}
\vspace{1.cm}
{\bf Table 2}\\
The dimensionless mass $\mu(\infty)$
of the first few SU(4) EYM black hole solutions
of the sequences corresponding to
case 1a of Table~1,
case 2a with node structure $(n,0)$ and
case 2b with node structure $n$ and $0$ is presented
for their respective extremal horizons
and for the horizon radius $x_{\rm H}=\sqrt{10}$.
For each sequence the corresponding limiting value of the mass
is shown in the last row (denoted by $\infty$).

\end{table}
\newpage
\begin{table}[p!]
\begin{center}
\begin{tabular}{|c|c|c|c|c|c|c|c|} \hline
\multicolumn{1}{|c|} { $ $ }&
\multicolumn{7}{ c|} {  $\mu(\infty)$ }\\
\hline
 & &  & &\multicolumn{2}{ c|} {n,0,n}     &   & \\
\cline{5-6}
\raisebox{1.5ex}[-1.5ex]{$n$} &  \raisebox{1.5ex}[-1.5ex]{$n,0,0$} &
\raisebox{1.5ex}[-1.5ex] {$0,n,0$} & \raisebox{1.5ex}[-1.5ex]{n,n,0} &
$u_1 = u_3 \ne u_2$ & $u_1 \ne u_3 \ne u_2$ &
\raisebox{1.5ex}[-1.5ex] {n+1,1,0} &\raisebox{1.5ex}[-1.5ex] {n+1,0,1} \\
\hline    
$1$      & 1.72400 & 2.01903 & 2.44237 & 2.32822 & 2.29105 & 2.70316 & 2.66330\\
$2$      & 2.21466 & 2.57219 & 2.86605 & 2.82600 & 2.82138 & 2.79238 & 2.78119\\
$3$      & 2.37577 & 2.74891 & 2.96878 & 2.95668 & 2.95588 & 2.82453 & 2.82109\\
$4$      & 2.42674 & 2.80401 & 2.99439 & 2.98933 & 2.98916 & 2.83425 & 2.83323\\
$5$      & 2.44252 & 2.82096 & 2.99879 & 2.99738 & 2.99734 & 2.83740 & 2.83707\\
$\infty$ & 2.44949 & 2.82843 & 3.0     & 3.0     & 3.0     & 2.83874 & 2.83874\\
 \hline
\end{tabular}
\end{center}
\vspace{1.cm}
{\bf Table 3
}\\
The dimensionless mass $\mu(\infty)$ of the first few solutions
of several globally regular SU(4) EYM sequences is presented.
For each sequence the corresponding limiting value of the mass
is shown in the last row (denoted by $\infty$).

\end{table}

\newpage
\begin{table}[p!]
\begin{center}
\begin{tabular}{|c|c|c|c|c|c|c|} \hline
  \multicolumn{2}{|r|} {n} & 1 & 2 & 3 & 4 & 5 \\
\hline
 & {$x_{\rm H \, ,n }^{\rm cr,1}$} & 4.1575 & 2.9564 & 2.8362 & 2.8165 & 2.8133\\
 \cline{2-7} 
 &{$\mu_{\rm reg}(\infty)$ } {(upper)}&2.7636 &3.0925 &3.1507 &3.1604 & 3.1620 \\
 \cline{2-7}
1 &{$\mu_{\rm reg}(\infty)$ } {(lower)}&2.7394 &3.0887 &3.1501 &3.1603 & 3.1620 \\
\cline{2-7}
 & {$\mu_{x_{\rm H}^{\rm cr,1}}(\infty)$} &3.1874 &3.1322 &3.1568 &3.1614 & 3.1621 \\
\hline 
 \hline
 &  {$x_{\rm H \, ,n}^{\rm cr,2}$} & 2.7143 & 2.1738 & 2.0904 & 2.0772 & 2.0741 \\
 \cline{2-7}
2 &{$\mu_{\rm reg}(\infty)$ } & 2.6734 & 3.0780 & 3.1484 & 3.1600 & 3.1619  \\
\cline{2-7}
 &  {$\mu_{x_{\rm H}^{\rm cr,2}}(\infty)$} & 2.8880 & 3.1059 & 3.1528 & 3.1607 & 3.1620\\
\hline
\hline
 &  {$x_{\rm H \, ,n}^{\rm cr,3}$} & 1.0246 & 0.5595 & 0.4741 &  &  \\
 \cline{2-7} 
 3 &{$\mu_{\rm reg}(\infty)$ } &2.7436 & 3.0889&3.1502 &  &  \\
\cline{2-7}
 &  {$\mu_{x_{\rm H}^{\rm cr,3}}(\infty)$} &2.7816 & 3.0925 &3.1507 &  &  \\
\hline
\hline
\multicolumn{2}{|c|} {$\mu_{\rm reg}(\infty)$ ($u_1=u_3=u_2$) } & 2.6204 & 3.0717 & 3.1475 & 3.1599 & 3.1619 \\ 
\hline
\end{tabular}
\end{center}
{\bf Table 4
}\\
The bifurcation behaviour of the SU(4) EYM $(n,n,n)$ sequences is presented
for $n=1-5$.
The mass of the regular solutions is shown
together with the corresponding values of the critical horizon and the mass 
of the black hole solutions at the critical horizon \cite{foot4}. 
For the solutions with $u_1 = u_3 \ne u_2$ (first set), 
there are two non-degenerate branches.
For comparison, the mass of the scaled SU(2) solutions
with $u_1 = u_3 = u_2$ is also presented. 

\end{table}
\newpage
\begin{table}[p!]
\begin{center}
\begin{tabular}{|c|r|c|c|c|c|} \hline
\multicolumn{2}{|c|}{ $(n,0,n)$ } &
\multicolumn{4}{ c|} { $x_{\rm H}=0$ } \\
\hline
\multicolumn{2}{|c|}{}&
\multicolumn{2}{ c|} {$\mu(\infty)$ }&
\multicolumn{2}{ c|} { $T/T_S$ }\\
\hline
$n$ & $\gamma =$ & 0 & 1 & 0 & 1  \\
 \hline 
  & $u_1\ne u_3$ & 2.3282 & 1.6269 & 0.1031 & 0.3445 \\
   \cline{2-6}
\raisebox{1.5ex}[-1.5ex]{1} & $u_1 = u_3$ &  2.2910 & 1.5919 & 0.1559 & 0.4296 \\
 \hline 
  & $u_1\ne u_3$ & 2.9567 & 2.0886 & 0.0075 & 0.0954 \\  
 \cline{2-6}
\raisebox{1.5ex}[-1.5ex]{3} & $u_1 = u_3$ & 2.9559 & 2.0875 & 0.0094 & 0.1104 \\
 \hline 
  & $u_1\ne u_3$ & 2.9974 & 2.1193 & 0.0005 & 0.0237 \\  
 \cline{2-6}
\raisebox{1.5ex}[-1.5ex]{5} & $u_1 = u_3$ & 2.9973 & 2.1193 & 0.0006 & 0.0272 \\
 \hline 
 {$\infty$} &  & 3.0 & 2.1215 & 0.0  & 0.0 \\  
 \cline{2-6}
 \hline 
\end{tabular}
\vspace{1cm}

\begin{tabular}{|cr|c|c|c|c|c|c|c|c|} \hline
\multicolumn{2}{|c|} {$(n,0,n)$ } &
\multicolumn{2}{ c|} {$x_{\rm H}^{\rm cr}$ } &
\multicolumn{2}{ c|} {$\mu(\infty)$ }&
\multicolumn{2}{ c|} { $T/T_S$ }&
\multicolumn{2}{ c|} { $u_1(x_{\rm H}^{\rm cr})$ }\\
\hline
$n$ & $\gamma =$ & 0 & 1 & 0 & 1 & 0 & 1 & 0 & 1 \\
\hline
1 & & 1.9723 & 1.7742 & 2.5065 & 2.0424 & 0.3038 & 0.5860 & 0.8632 & 0.8552 \\  
 \hline
3 & & 1.2440 & 0.2927 & 2.9633 & 2.1058 & 0.0154 & 0.1411 & 0.9063 & 0.8812 \\
 \hline
5 & & 1.1803 & 0.0690 & 2.9978 & 2.1203 & 0.0009 & 0.0345 & 0.9130 & 0.8860 \\
\hline
\end{tabular}
\end{center}
{\bf Table 5
}\\
{\sl Upper part:}
The dimensionless mass $\mu(\infty)$
of the first few $u_1 \ne u_3$ and $u_1 = u_3$
solutions with odd $n$ is shown for
the two globally regular sequences with node structure $(n,0,n)$ 
for SU(4) EYM theory and SU(4) EYMD theory with $\gamma=1$.
Also shown is the temperature $T/T_S$
($T_S=(4 \pi x_{\rm H})^{-1}$),
to be interpreted
as the $x_{\rm H} \rightarrow 0$ limit of the black hole solutions.
The corresponding limiting values
are shown in the last row (denoted by $\infty$).

{\sl Lower part:}
The critical values of the horizon radius $x_{\rm H \, n}^{\rm cr}$,
where the $u_1 \ne u_3$ solutions with node structure $(n,0,n)$
merge into the corresponding $u_1 = u_3$ solutions
are shown for SU(4) EYM theory and SU(4) EYMD theory with $\gamma=1$.
Also shown are the dimensionless mass $\mu(\infty)$,
the temperature $T/T_S$ and 
$u_1(x_{\rm H }^{\rm cr})$.

\end{table}
\newpage
\noindent
\begin{figure}
\centering
\epsfysize=11cm
\mbox{\epsffile{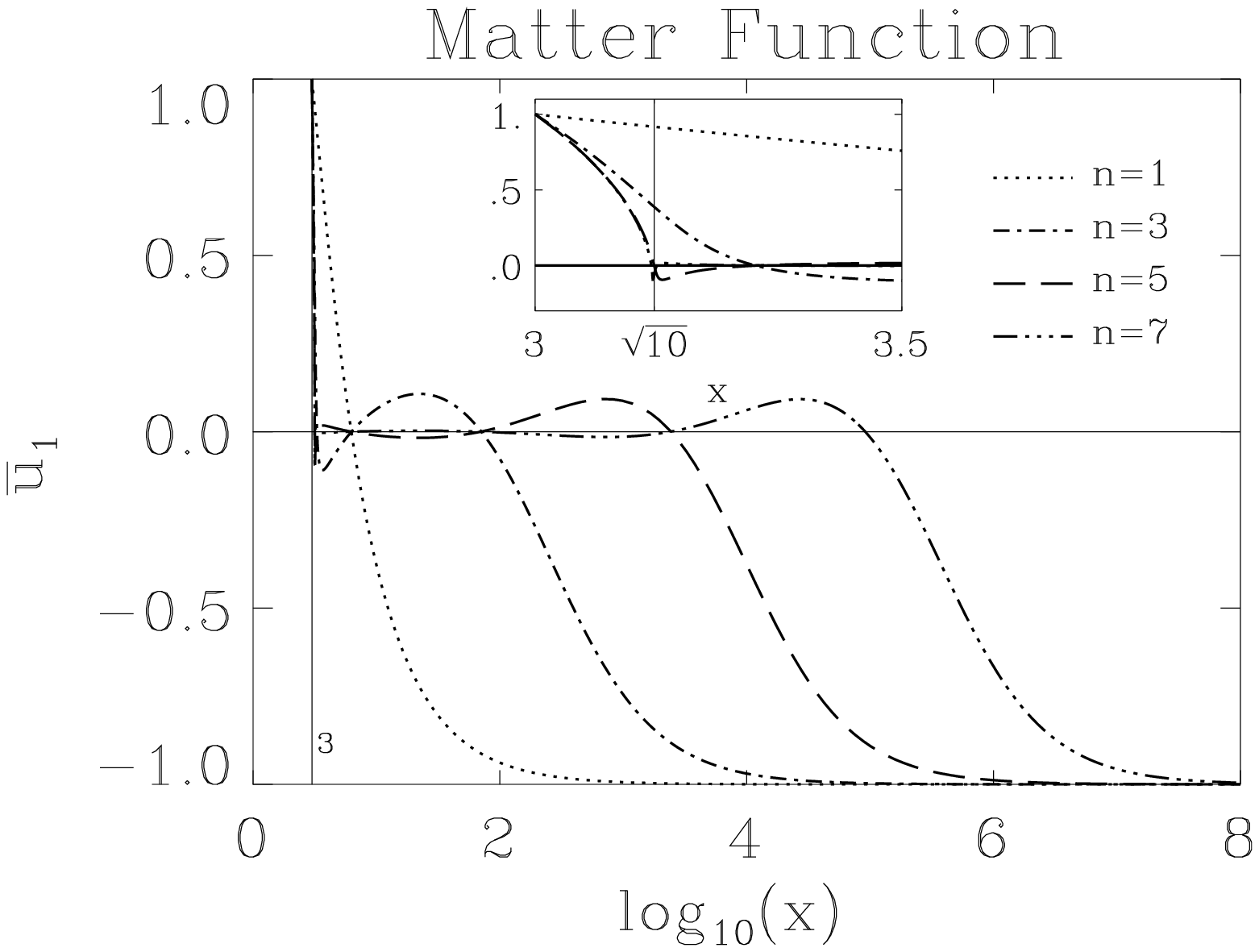
}}
\end{figure}

Fig.~1a: The gauge field function $\bar u_{1}(x)$ 
is shown as a function of the dimensionless coordinate $x$
for the charged SU(4) EYM black hole solutions of case 1a of Table~1
with extremal event horizon $x_{\rm H} = 3$
and node numbers $n=1$, 3, 5, 7.
The inset illustrates the convergence of the innermost node
to a value just below $x = \sqrt{10}$ 
(thin vertical line) in the limit $n \rightarrow \infty$.
The thin horizontal line represents the limiting function for 
$x> \sqrt{10}$.

\newpage
\begin{figure}
\centering
\epsfysize=11cm
\mbox{\epsffile{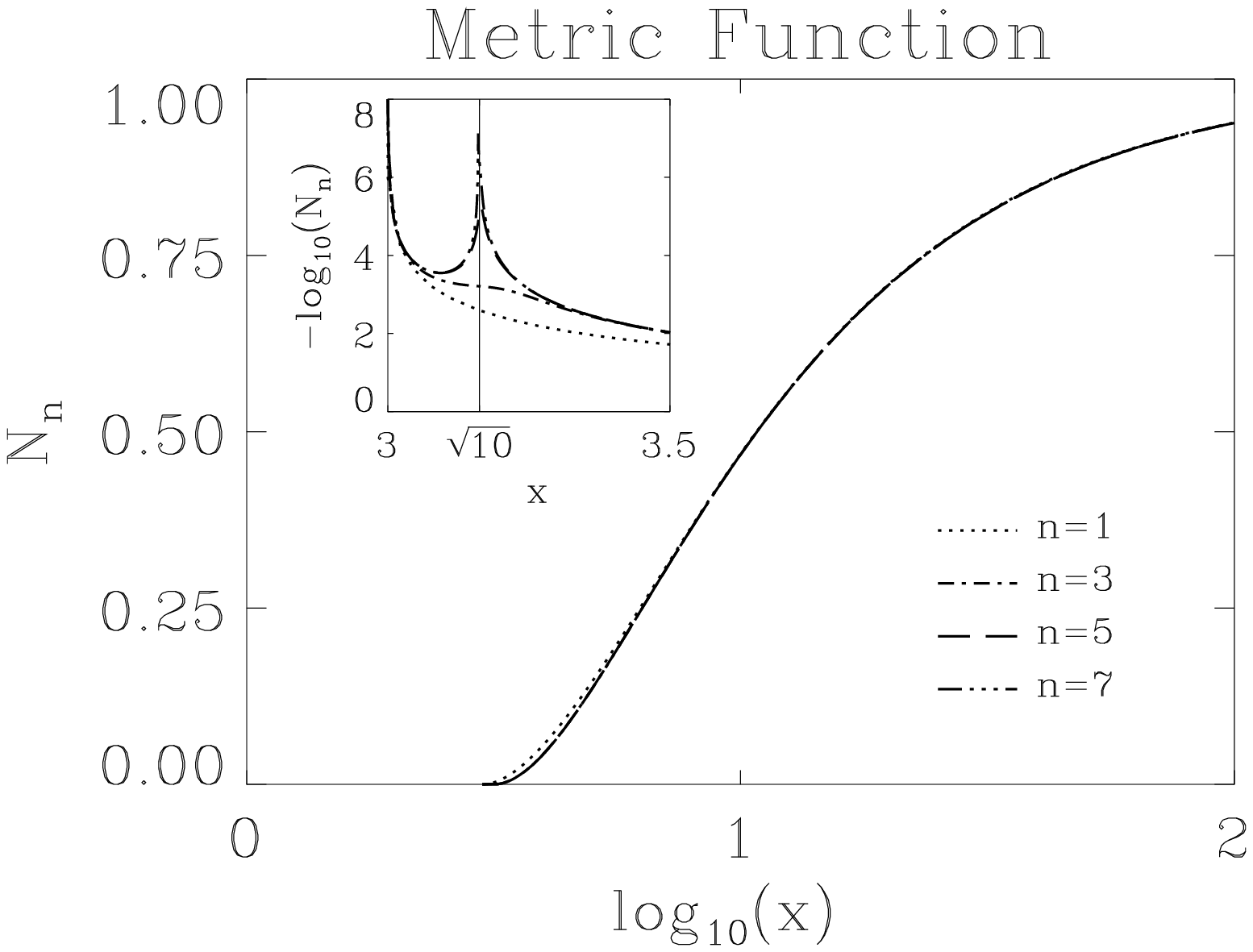
}}
\end{figure}

Fig.~1b: Same as Fig.~1a for the metric function ${\cal N}(x)$.
The inset illustrates the emergence of a second zero at
$x = \sqrt{10}$ in the limit $n \rightarrow \infty$.

\newpage
\begin{figure}
\centering
\epsfysize=11cm
\mbox{\epsffile{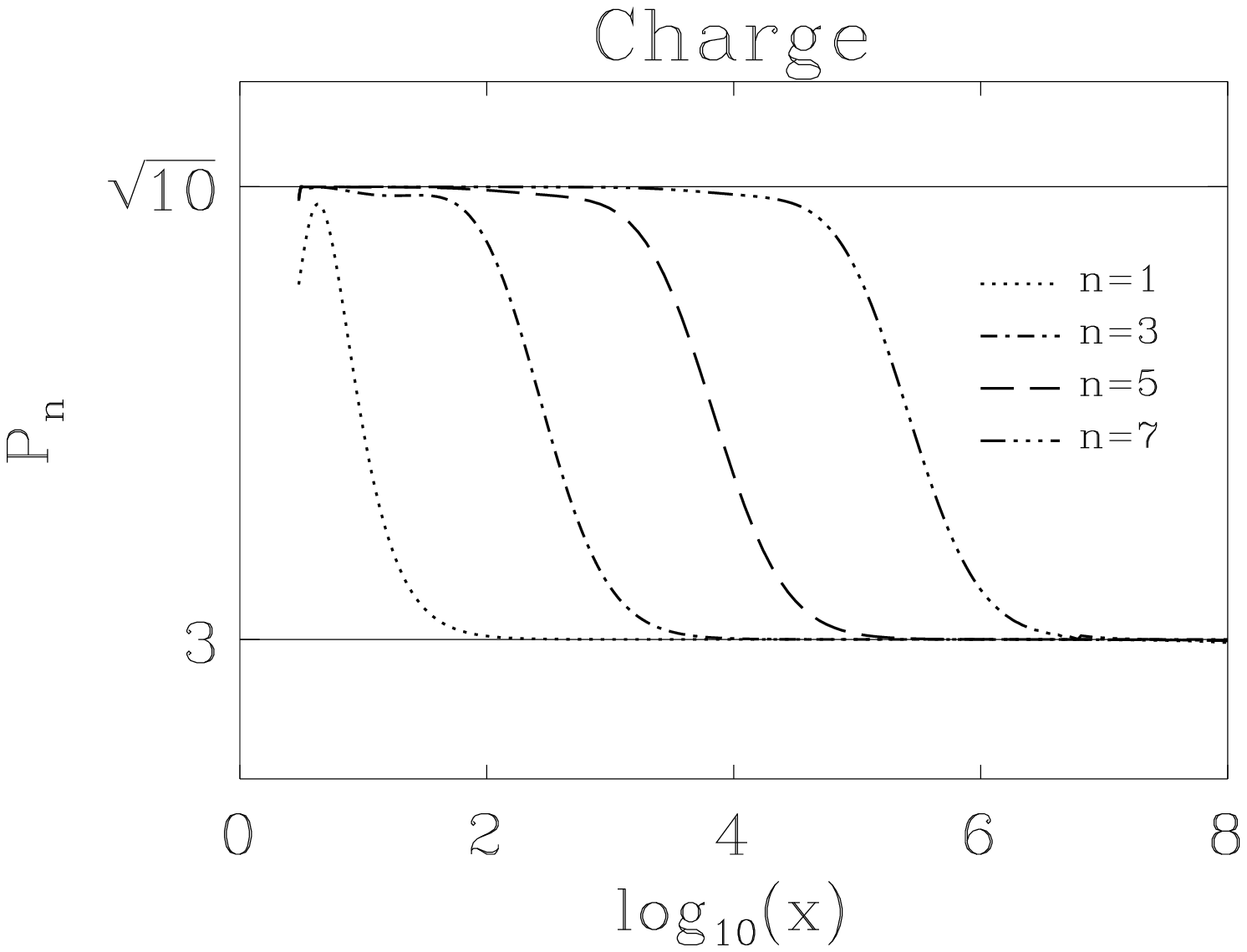
}}
\end{figure}

Fig.~1c: 
Same as Fig.~1a for the charge function $P(x)$.
The thin horizontal line represents the norm of the charge of the
limiting solution, $P = \sqrt{10}$.

\newpage
\begin{figure}
\centering
\epsfysize=11cm
\mbox{\epsffile{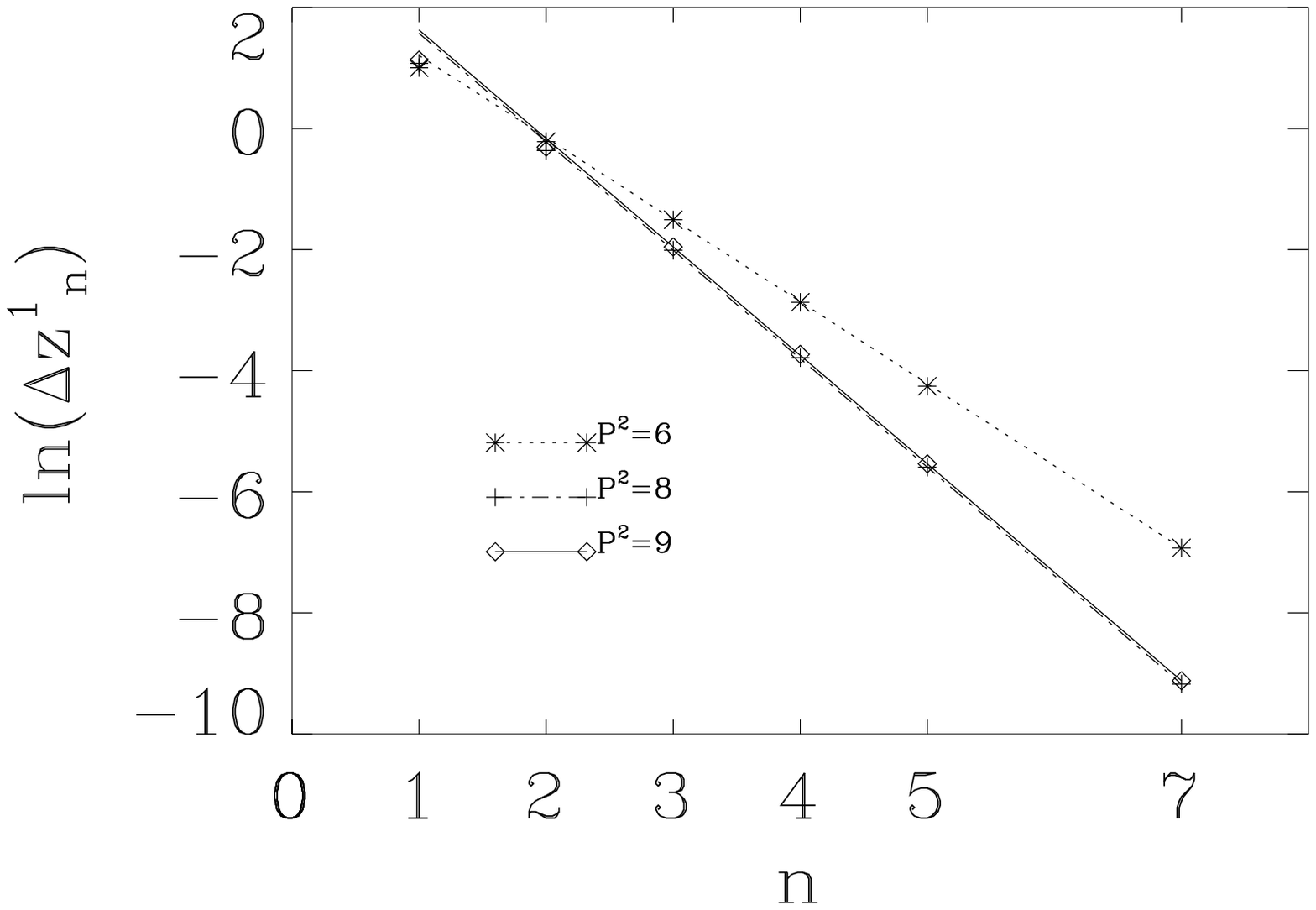
}}
\end{figure}

Fig.~2: 
The logarithm of the absolute deviation from the limiting solution 
$\Delta z^1_n = |z^1_\infty -z^1_n| $ is shown
as a function of the node number $n$ 
for the location of the innermost node 
of the extremal SU(4) EYM black hole solutions
of cases 1a, 2a and 2b of Table~1.

\newpage
\begin{figure}
\centering
\epsfysize=11cm
\mbox{\epsffile{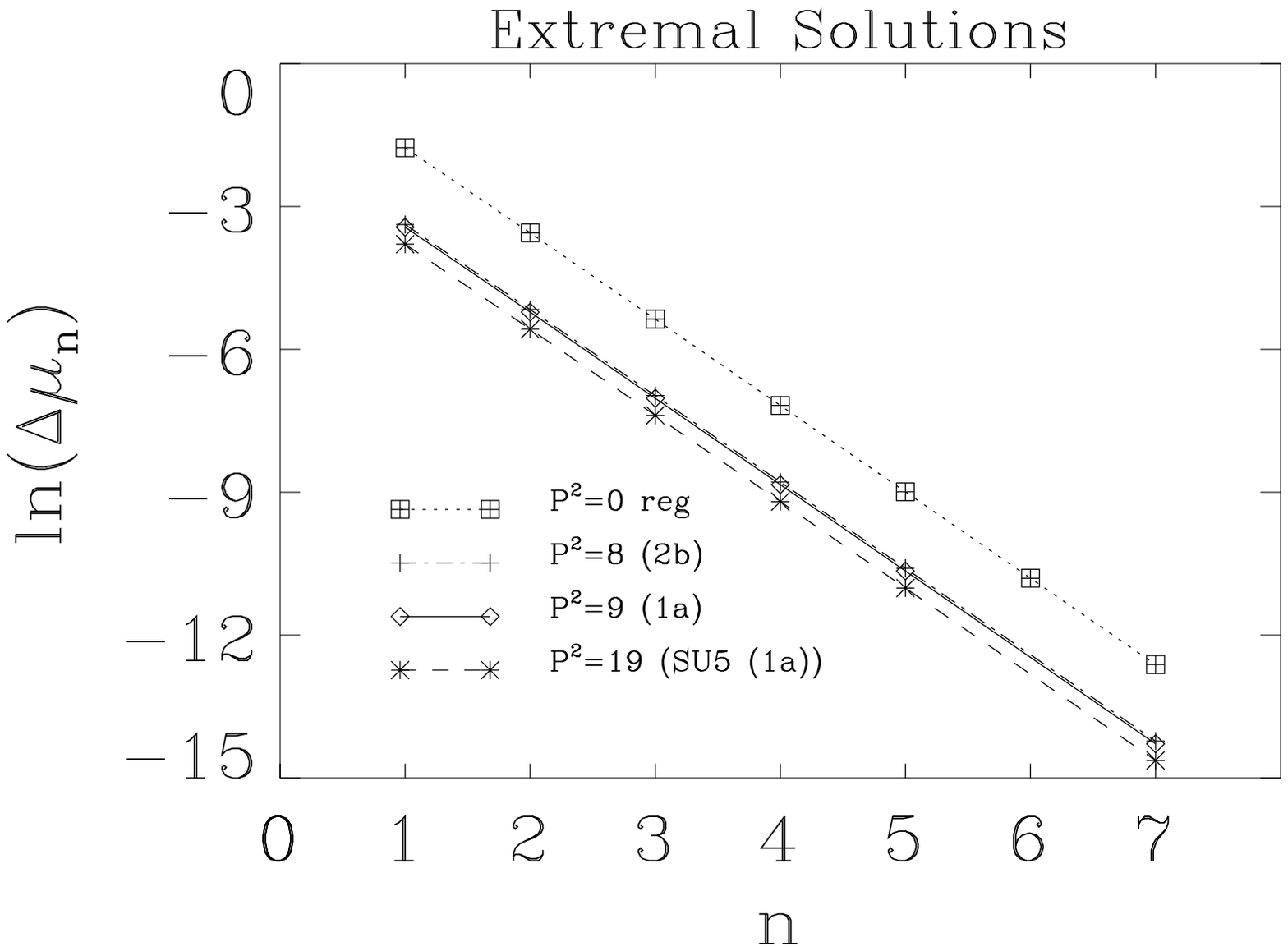
}}
\end{figure}

Fig.~3: 
The logarithm of the absolute deviation from the limiting solution 
$\Delta \mu_n = \mu_\infty (\infty)-\mu_n (\infty)$ is shown
as a function of the node number $n$ for the mass
of the extremal SU(4) EYM black hole solutions
of cases 1a and 2b of Table~1, together with 
case 1a of ref.~\cite{kks5}, representing
extremal SU(5) EYM black hole solutions,
as well as for the globally regular SU(2) EYM solutions.

\newpage
\begin{figure}
\centering
\epsfysize=11cm
\mbox{\epsffile{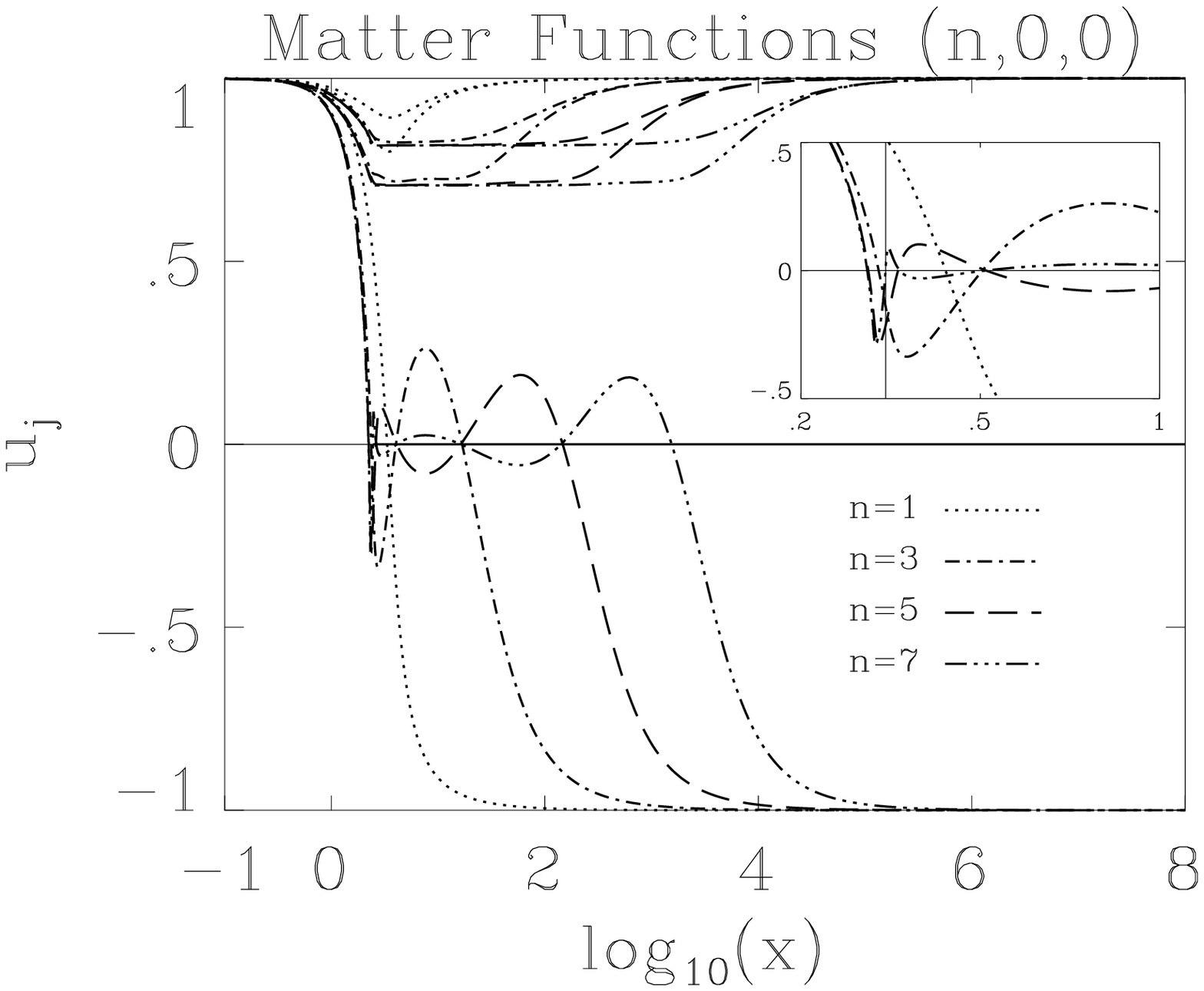
}}
\end{figure}

Fig.~4a: 
The gauge field functions $u_{j}(x)$, $j=1-3$,
are shown as functions of the dimensionless coordinate $x$
for the globally regular SU(4) EYM solutions with node structure $(n,0,0)$
and node numbers $n=1$, 3, 5, 7.
The inset illustrates the convergence of the innermost node
to a value just below $x=\sqrt{6}$
(thin vertical line) in the limit $n \rightarrow \infty$.
The thin horizontal line represents the limiting function for 
$x> \sqrt{6}$.

\newpage
\begin{figure}
\centering
\epsfysize=11cm
\mbox{\epsffile{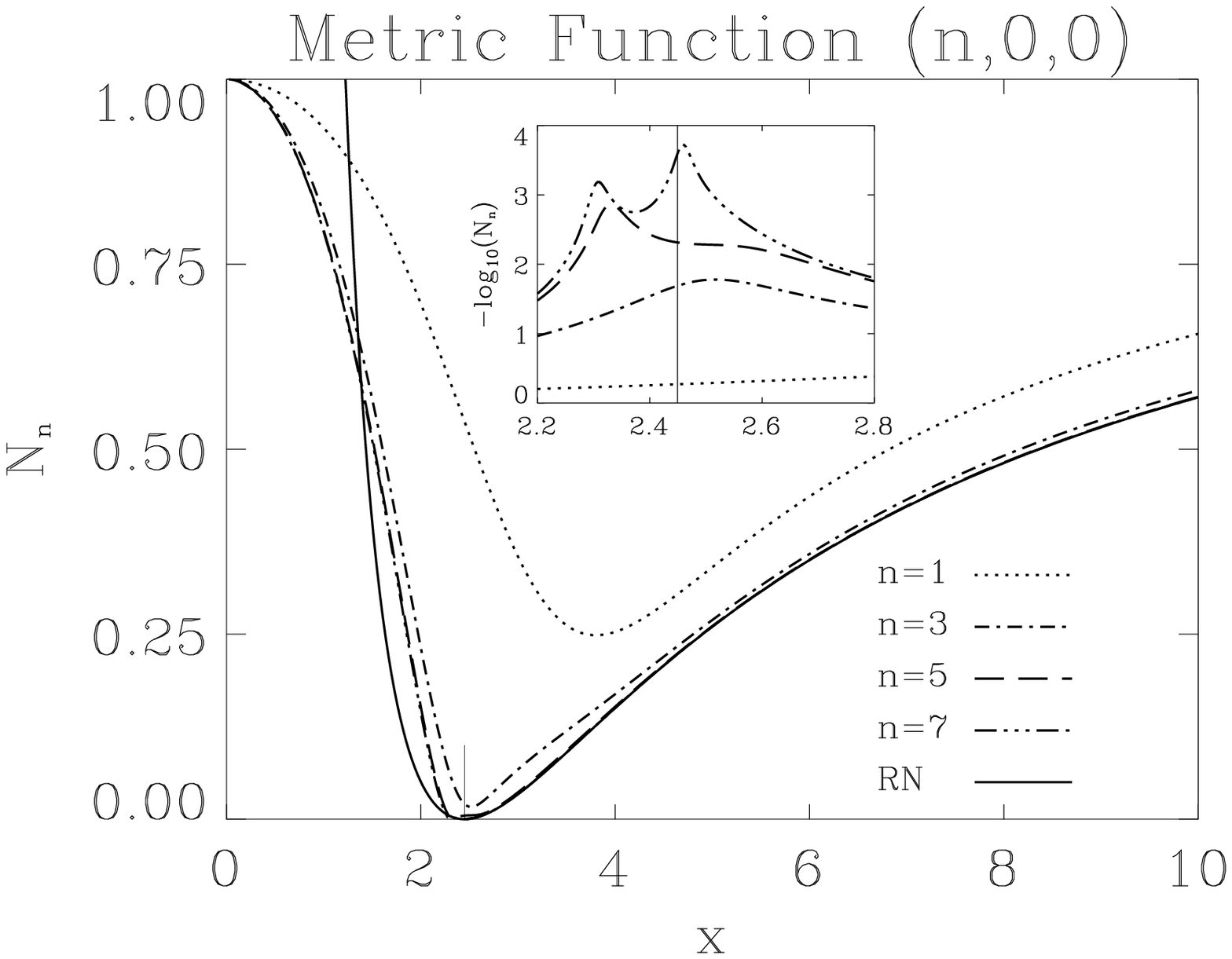
}}
\end{figure}

Fig.~4b:
Same as Fig.~4a for the metric function ${\cal N}(x)$.
The inset illustrates the emergence of a second zero at
$x=\sqrt{6}$ in the limit $n \rightarrow \infty$.

\newpage
\begin{figure}
\centering
\epsfysize=11cm
\mbox{\epsffile{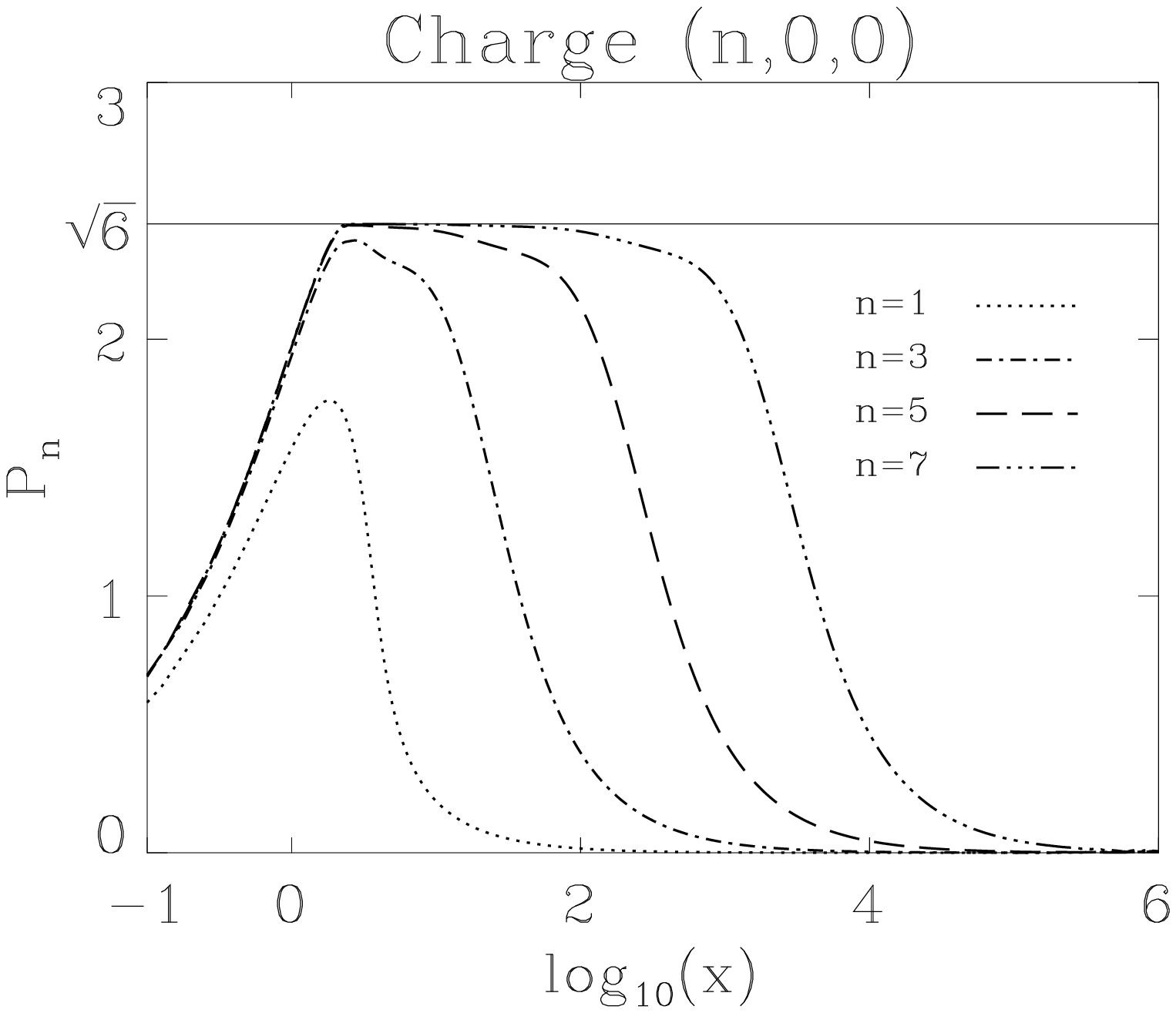
}}
\end{figure}

Fig.~4c: 
Same as Fig.~4a for the charge function $P(x)$.
The thin horizontal line represents the norm of the charge of the
limiting solution, $P = \sqrt{6}$.

\newpage
\begin{figure}
\centering
\epsfysize=11cm
\mbox{\epsffile{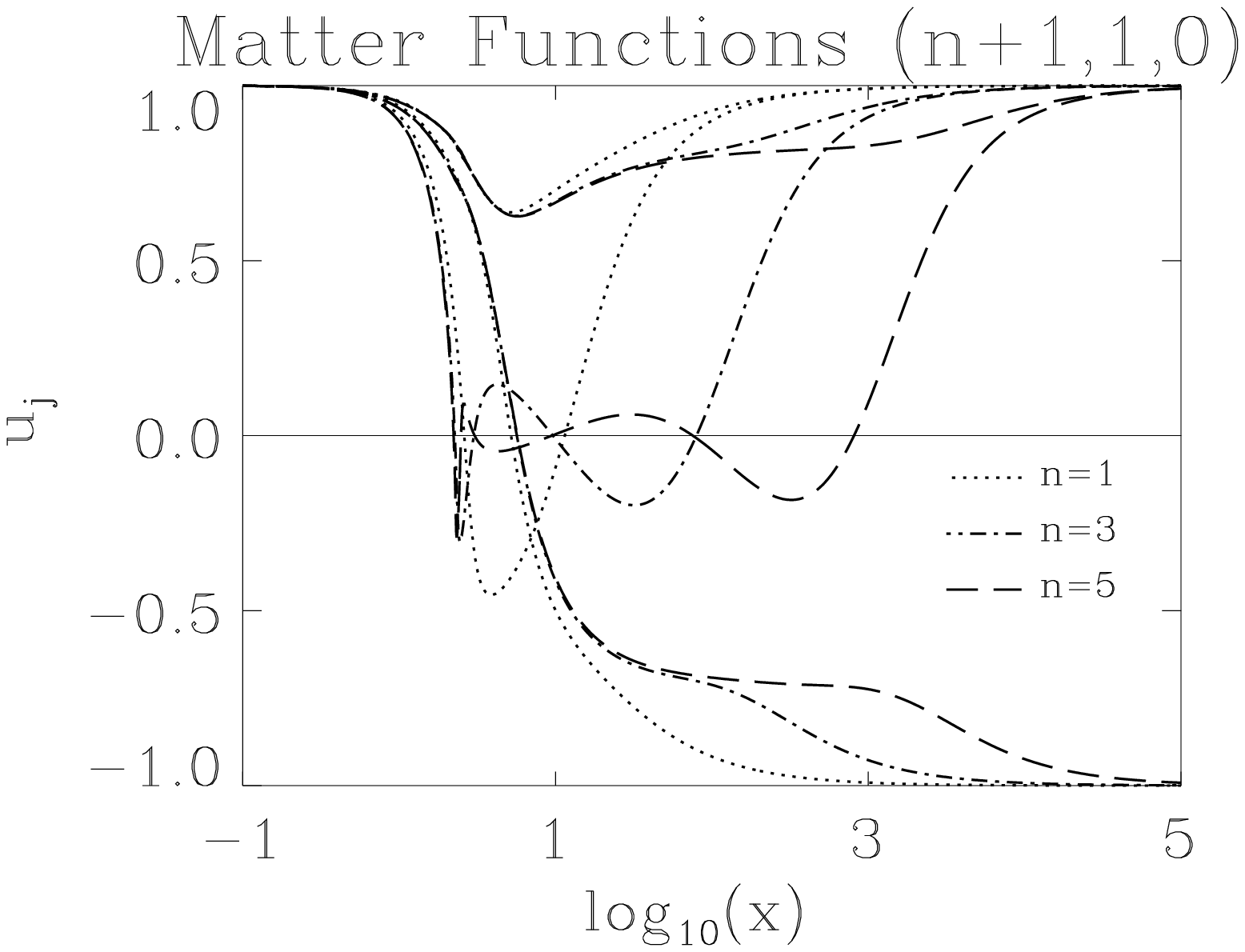
}}
\end{figure}

Fig.~5a:
The gauge field functions $u_{j}(x)$, $j=1-3$, 
are shown as functions of the dimensionless coordinate $x$
for the globally regular SU(4) EYM solutions with node structure $(n+1,1,0)$
and node numbers $n=1$, 3, 5.
The thin horizontal line represents the limiting function for 
$x> \sqrt{6}$.

\newpage
\begin{figure}
\centering
\epsfysize=11cm
\mbox{\epsffile{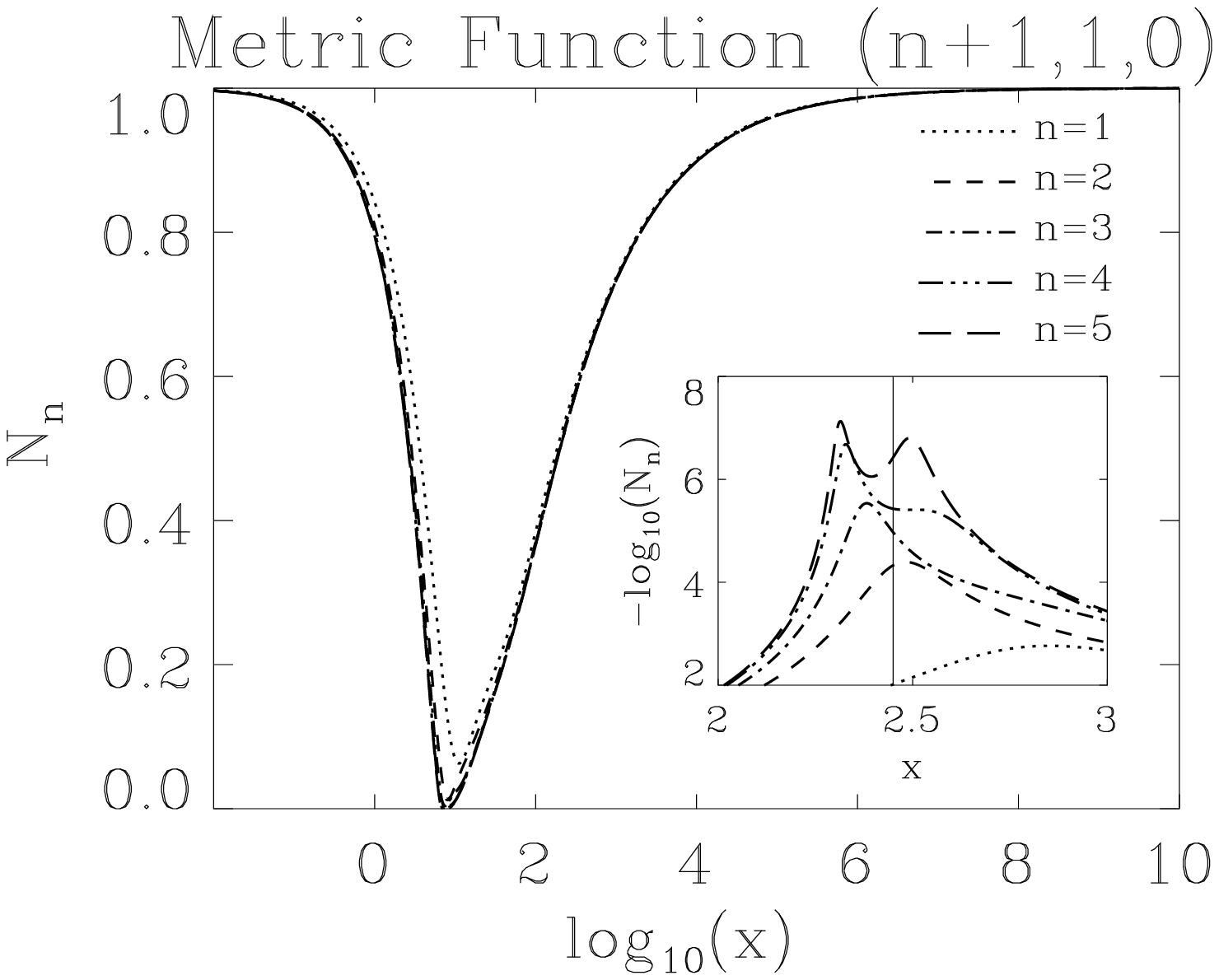
}}
\end{figure}

Fig.~5b: 
Same as Fig.~5a for the metric function ${\cal N}(x)$
and node numbers $n=1-5$.
The inset illustrates the emergence of a second zero at
$x=\sqrt{6}$ (thin vertical line) in the limit $n \rightarrow \infty$.

\newpage
\begin{figure}
\centering
\epsfysize=11cm
\mbox{\epsffile{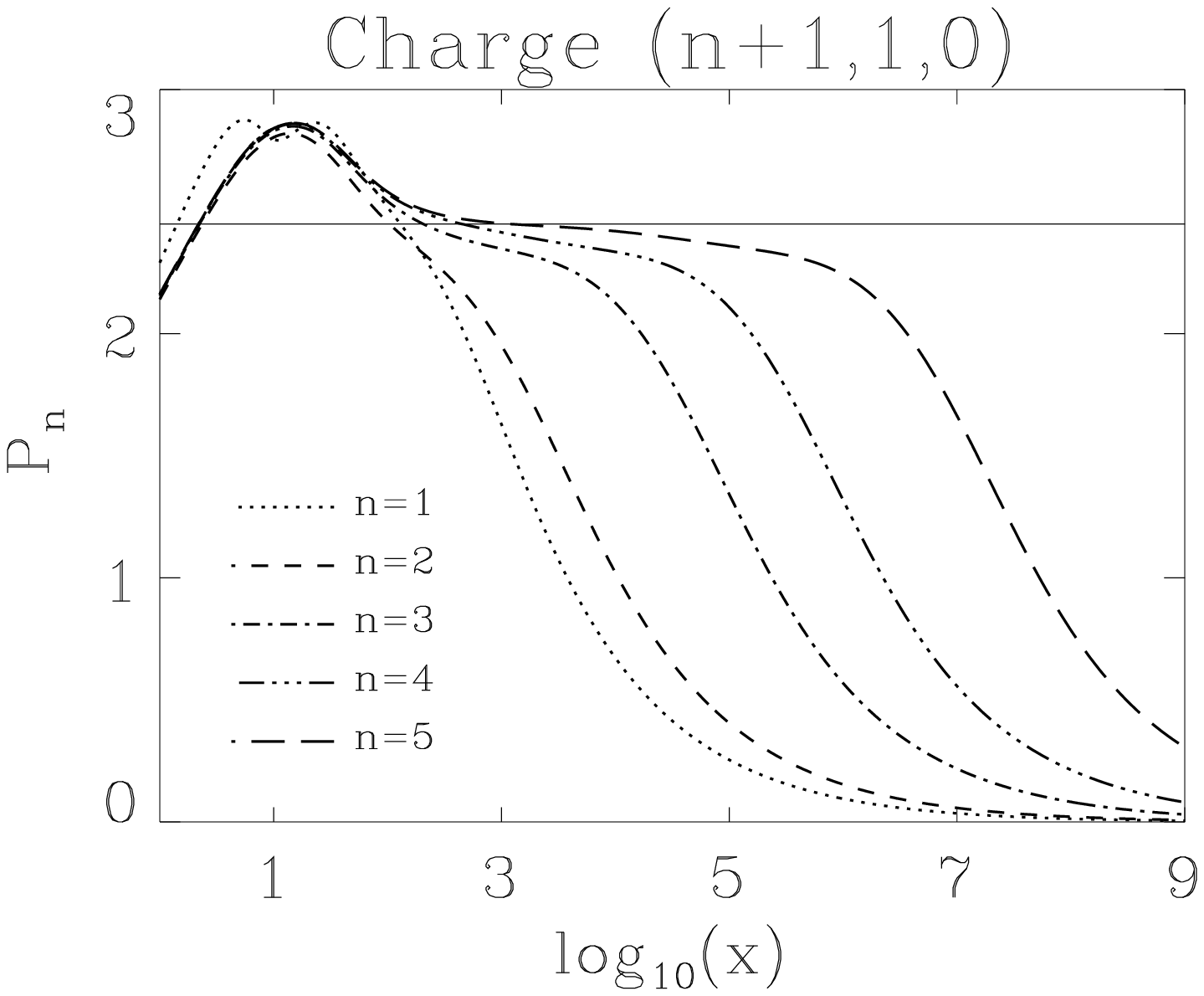
}}
\end{figure}

Fig.~5c:
Same as Fig.~5a for the charge function $P(x)$
and node numbers $n=1-5$.
The thin horizontal line represents the norm of the charge of the
limiting solution, $P = \sqrt{6}$.

\newpage
\begin{figure}
\centering
\epsfysize=11cm
\mbox{\epsffile{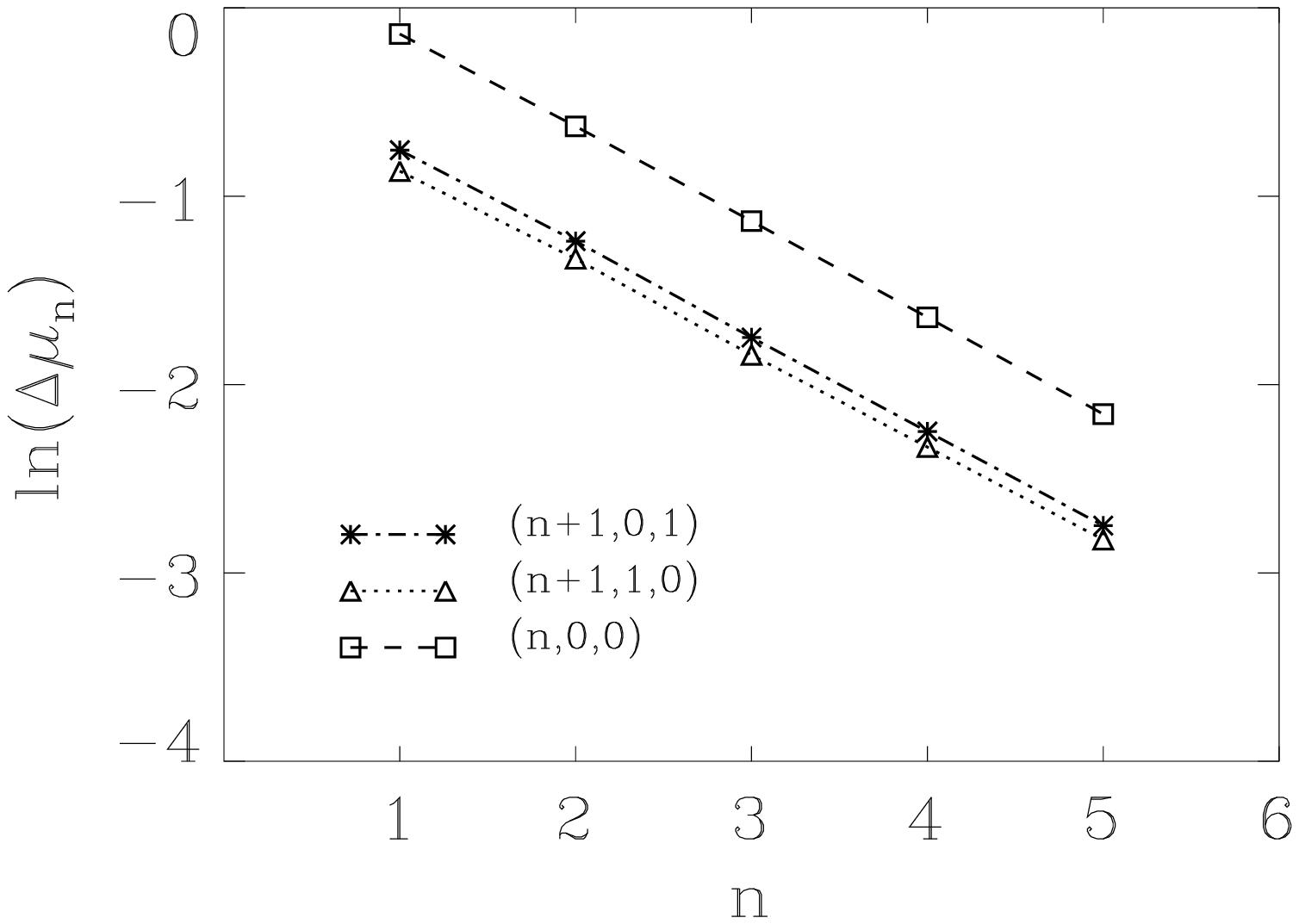
}}
\end{figure}

Fig.~6: 
The logarithm of the absolute deviation from the limiting solution 
$\Delta \mu_n = \mu_\infty (\infty)-\mu_n (\infty)$ is shown
as a function of the node number $n$ 
for the mass of the sequences of globally regular SU(4) EYM solutions
with node stucture $(n,0,0)$, $(n+1,1,0)$ and $(n+1,0,1)$.

\newpage
\begin{figure}
\centering
\epsfysize=11cm
\mbox{\epsffile{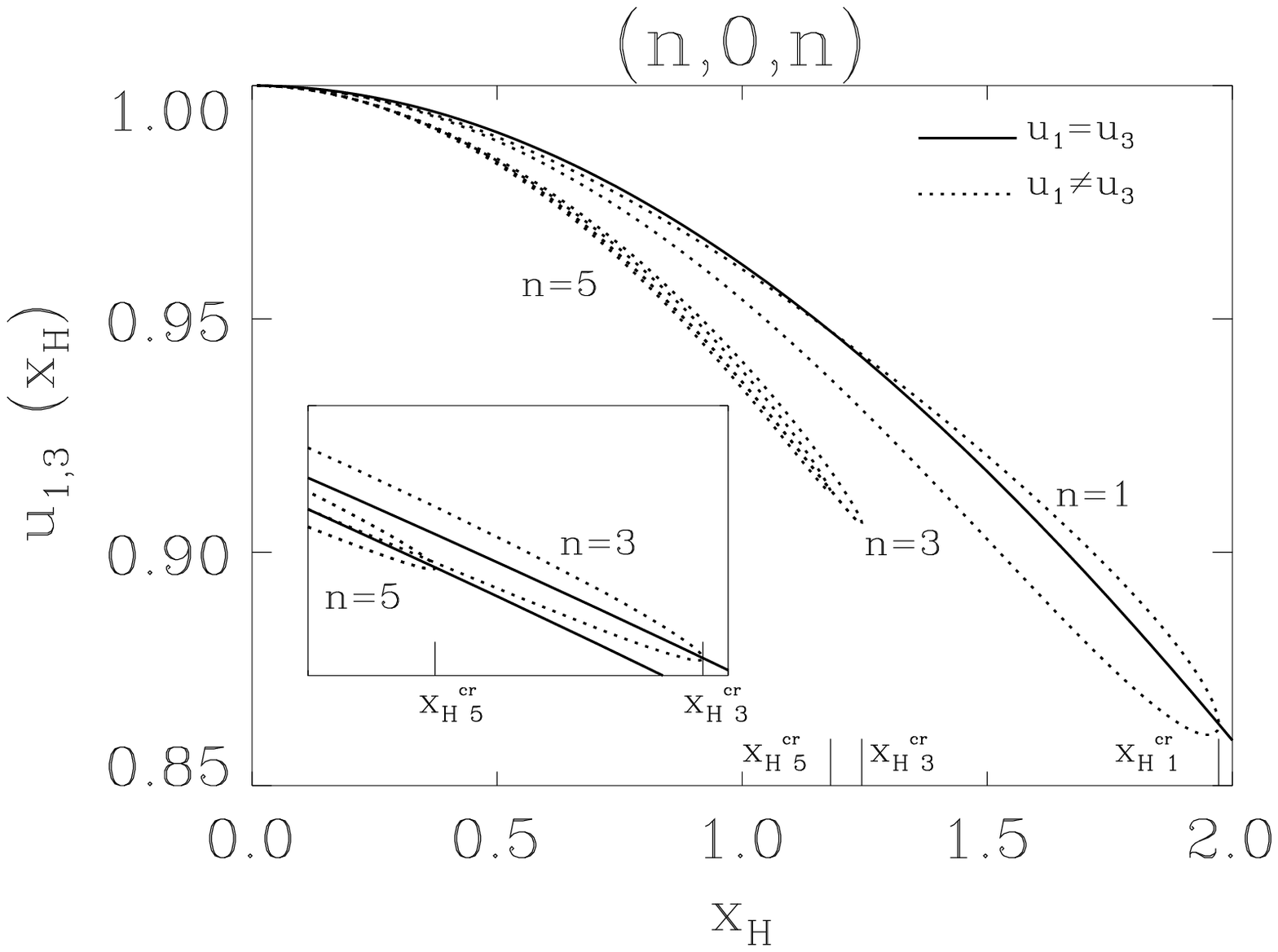
}}
\end{figure}

Fig.~7a: 
The value of the gauge field functions $u_1$ and $u_3$ at the horizon
is shown as a function of the horizon radius $x_{\rm H}$
for the SU(4) EYM black hole solutions with node structure $(n,0,n)$
and node numbers $n=1$, 3, 5.
For $n=1$ both $u_1 = u_3$ and $u_1 \ne u_3$ solutions are shown.
For $n=3$ and $n=5$ the $u_1 = u_3$ solutions are only
shown in the inset ($\Delta x_{\rm H}=0.1$, $\Delta u=0.02$).
The critical values of the horizon radius $x_{\rm H \, n}^{\rm cr}$
are also indicated.

\newpage
\begin{figure}
\centering
\epsfysize=11cm
\mbox{\epsffile{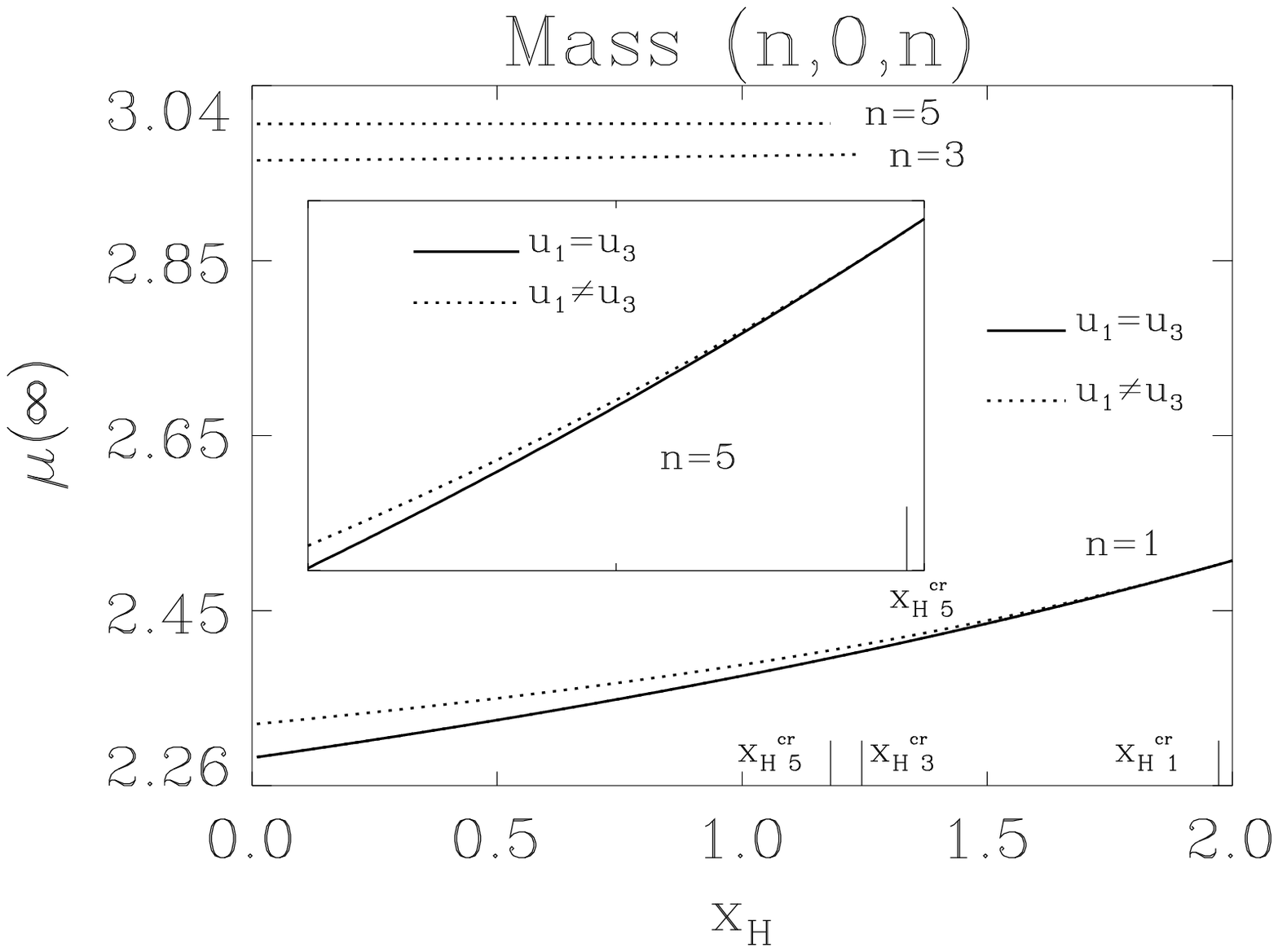
}}
\end{figure}

Fig.~7b: 
Same as Fig.~7a for the dimensionless mass $\mu(\infty)$.
The inset ($\Delta x_{\rm H}=0.7$, $\Delta \mu=0.0003$) only shows $n=5$.

\newpage
\begin{figure}
\centering
\epsfysize=11cm
\mbox{\epsffile{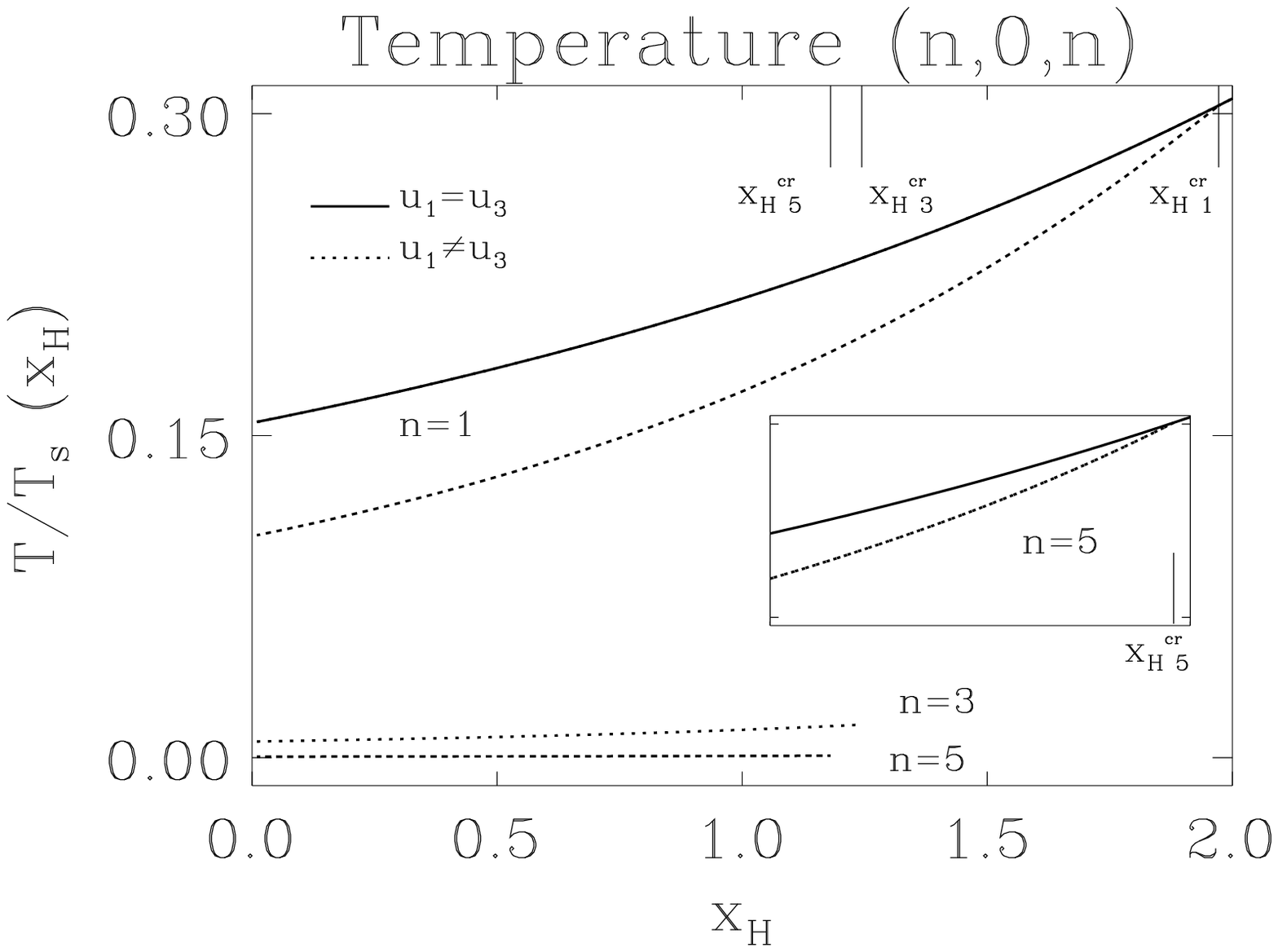
}}
\end{figure}

Fig.~7c: 
Same as Fig.~7a for the temperature $T/T_S$
($T_S={4 \pi x_{\rm H}}^{-1}$).
The inset ($\Delta x_{\rm H}=0.5$, $\Delta T/T_S=0.0003$) only shows $n=5$.

\newpage
\begin{figure}
\centering
\epsfysize=11cm
\mbox{\epsffile{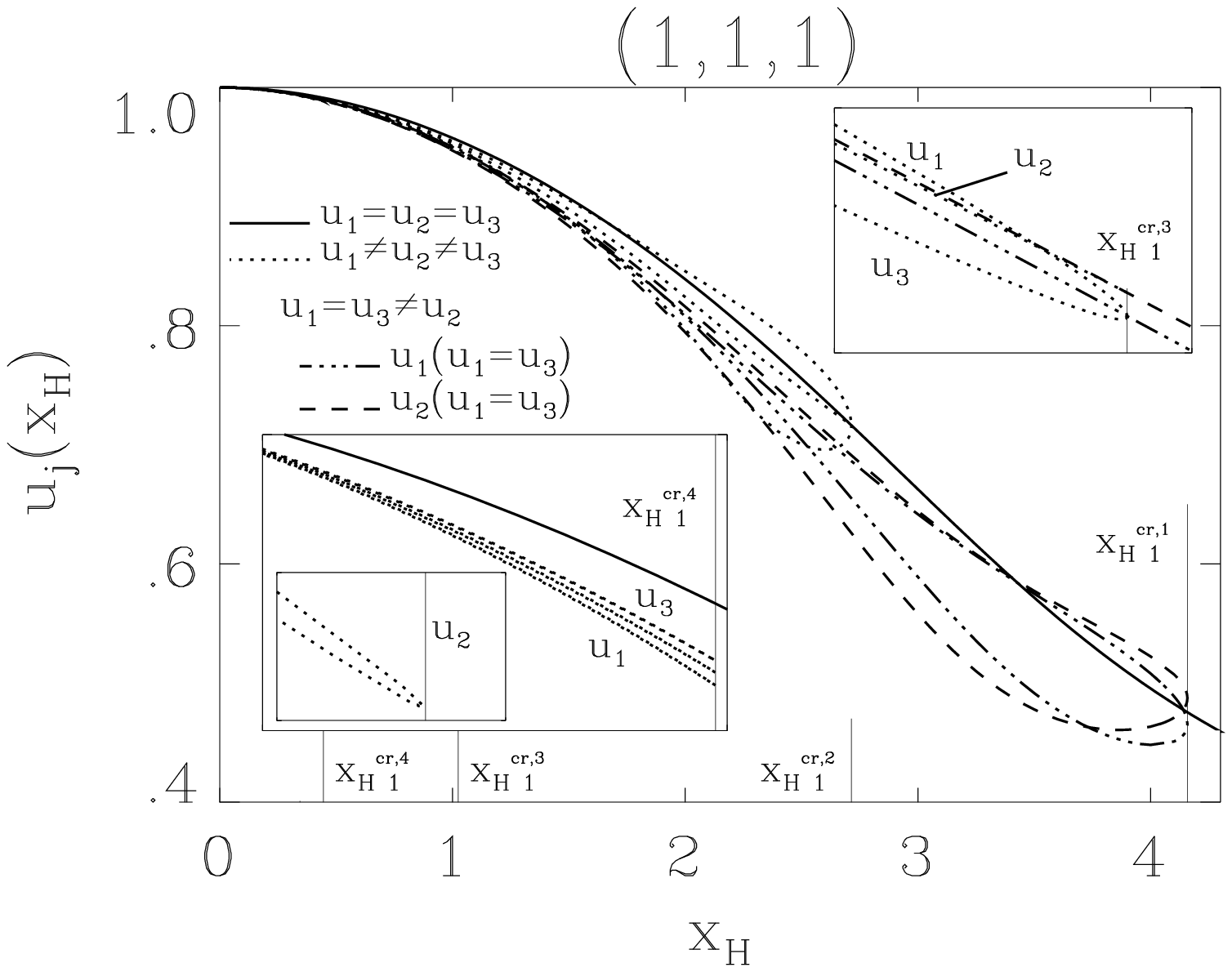
}}
\end{figure}

Fig.~8a:
The value of the gauge field functions $u_1$, $u_2$ and $u_3$ at
the horizon is shown as a function of the horizon radius $x_{\rm H}$
for the SU(4) EYM black hole solutions with node structure $(1,1,1)$.
The upper right inset ($\Delta x_{\rm H}=0.03$, $\Delta u=0.004$)
shows the merging of the degenerate branches of
$u_1 \ne u_3 \ne u_2$ solutions (dotted lines) 
into the lower branch of $u_1 = u_3 \ne u_2$ solutions at
$x_{\rm H \, 1}^{\rm cr,3}$.
The lower left inset ($\Delta x_{\rm H}=2 \times 10^{-4}$,
$\Delta u=9 \times 10^{-6}$)
shows two non-degenerate branches of $u_1 \ne u_3 \ne u_3$
solutions close to their critical value $x_{\rm H \, 1}^{\rm cr,4}$.

\newpage
\begin{figure}
\centering
\epsfysize=11cm
\mbox{\epsffile{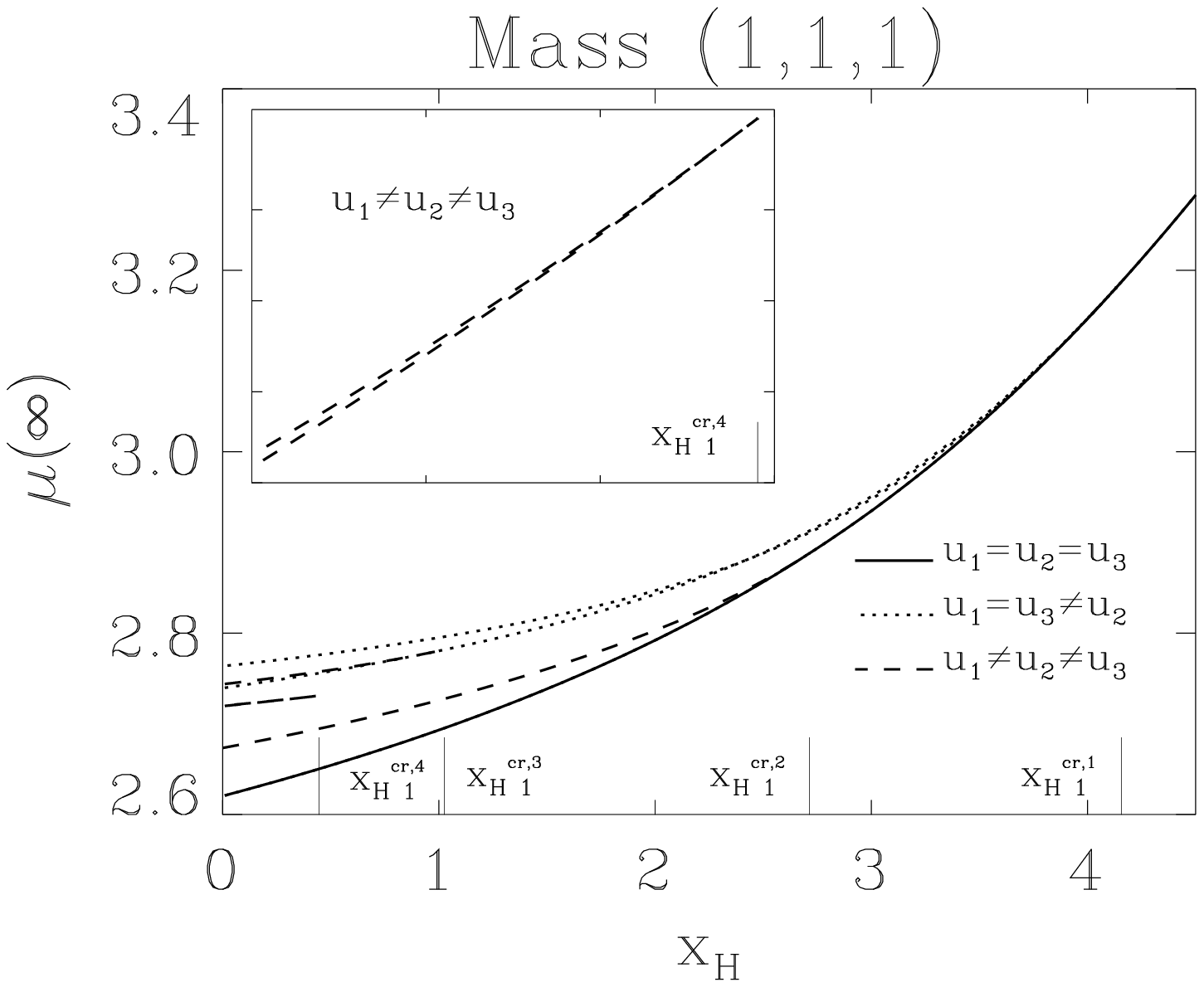
}}
\end{figure}

Fig.~8b:
Same as Fig.~8a for the dimensionless mass $\mu(\infty)$.
The inset ($\Delta x_{\rm H}=0.45$, $\Delta \mu=0.012$)
shows two non-degenerate branches of $u_1 \ne u_3 \ne u_2$ 
solutions close to their critical value $x_{\rm H \, 1}^{\rm cr,4}$.

\newpage
\begin{figure}
\centering
\epsfysize=11cm
\mbox{\epsffile{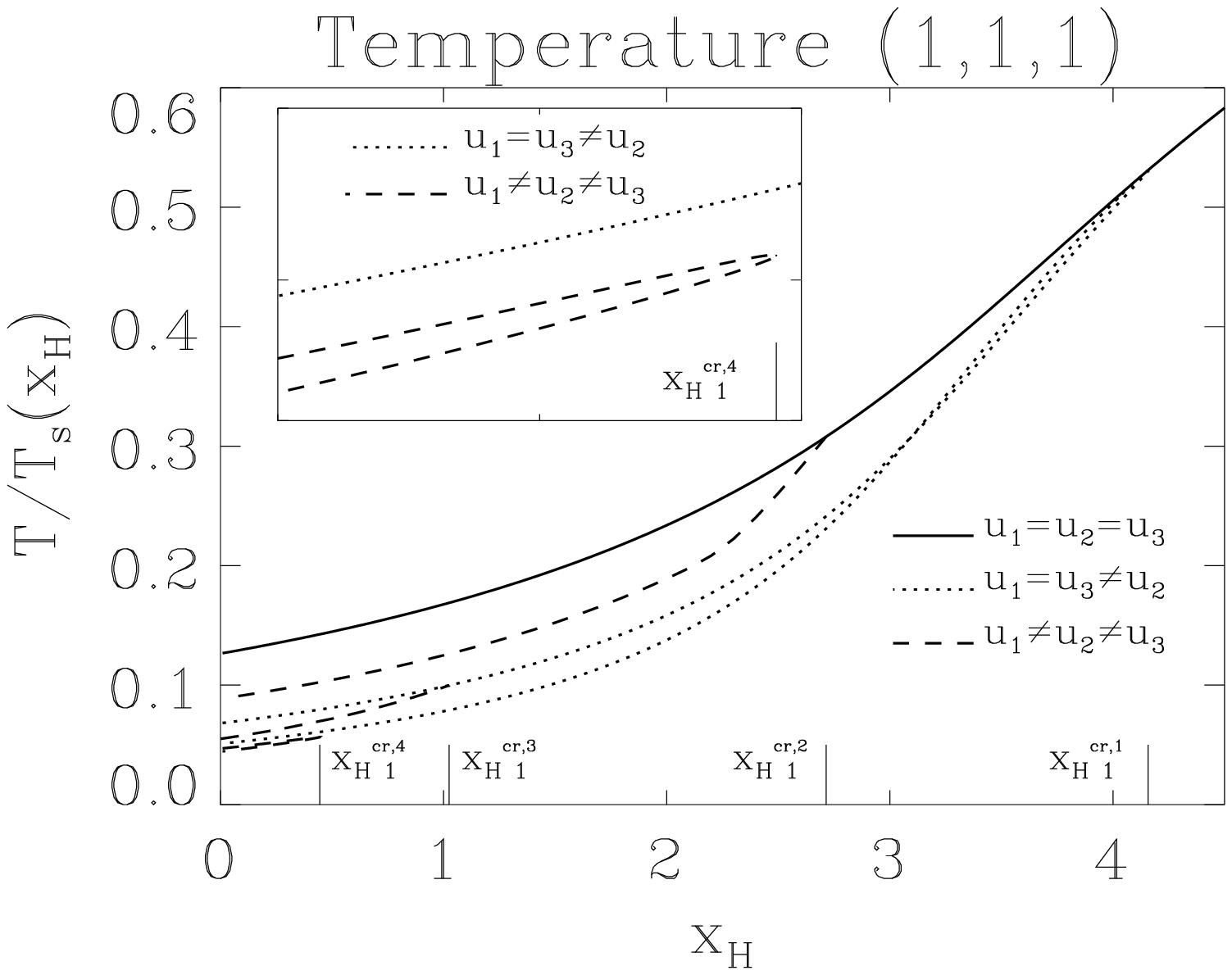
}}
\end{figure}

Fig.~8c:
Same as Fig.~8a for the temperature $T/T_S$.
The inset ($\Delta x_{\rm H}=0.3$, $\Delta T/T_S=0.02$)
shows two non-degenerate branches of $u_1 \ne u_3 \ne u_2$ 
solutions close to their critical value $x_{\rm H \, 1}^{\rm cr,4}$.

\newpage
\begin{figure}
\centering
\epsfysize=11cm
\mbox{\epsffile{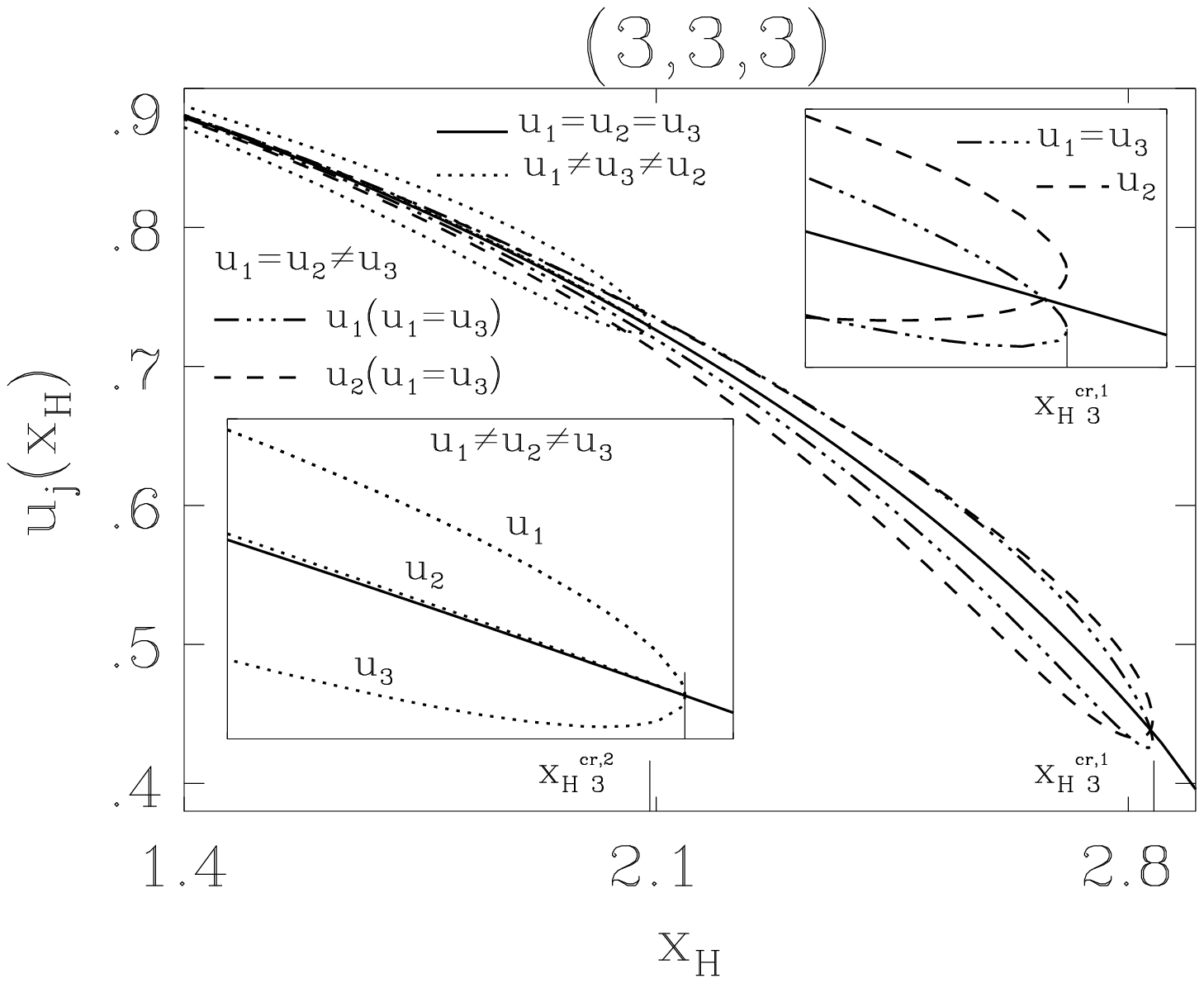
}}
\end{figure}

Fig.~9: 
Same as Fig.~8a for the
solution with node structure $(3,3,3)$.
The upper right inset ($\Delta x_{\rm H}=0.05$, $\Delta u=0.07$)
shows the two non-degenerate branches of 
$u_1 = u_3 \ne u_2$ solutions close to their critical value
$x_{\rm H \, 3}^{\rm cr,1}$ as well as the branch of scaled SU(2) solutions.
The lower left inset ($\Delta x_{\rm H}=0.1$, $\Delta u=0.05$)
shows two degenerate branches of $u_1 \ne u_3 \ne u_2$ solutions
close to their ciritical value
$x_{\rm H \, 3}^{\rm cr,2}$ as well as the branch of scaled SU(2) solutions.


\begin{thebibliography}{000}
\bibitem{foot1}
 In addition to the magnetic charge $P=\sqrt{3}$,
 SU(3) EYM black hole solutions can carry arbitrary electric charge $Q$,
 not considered here.
 In that case the event horizon of the extremal SU(3) EYM
 black hole solutions is determined by $x_{\rm H}=\sqrt{P^2+Q^2}$ \cite{gv}.
\bibitem{foot3}
 We expect (and find) only one type of solutions 
 with node structure $(n,n,0)$,
 because there is no symmetry with respect to $u_1 \leftrightarrow u_2$.
\bibitem{foot4}
 Solutions of the $(n,n,n)$ sequences with $u_1 \ne  u_3 \ne u_2$ 
 which merge into 
 the lower branch of solutions with $u_1 = u_3 \ne u_2$ at  
 $x_{\rm H \, n}^{\rm cr,3}$ have not been numerically constructed
 for $n=4$ or 5. Solutions corresponding to $x_{\rm H \, n}^{\rm cr,4}$
 have not been constructed for $n > 1$.
 Since the critical values of the horizon radius decrease with increasing $n$, 
 these may either be harder to obtain or 
 they might not exist any more.
\end{thebibliography}
\end{document}